\documentclass[12pt]{article}
\pdfoutput=1

\usepackage{draft} 
\usepackage[weather]{ifsym}

\usepackage{fontawesome}
\usepackage{hyperref}
\usepackage{graphicx,color,subfig}
\usepackage{cite}
\usepackage{mciteplus}
\usepackage{skak}
\usepackage{empheq}
\usepackage{tikz}
\usepackage{bbm}
\usepackage{url}

\DeclareFontFamily{OT1}{pzc}{}
\DeclareFontShape{OT1}{pzc}{m}{it}{<-> s * [1.10] pzcmi7t}{}
\DeclareMathAlphabet{\mathpzc}{OT1}{pzc}{m}{it}

\usetikzlibrary{snakes}
\usetikzlibrary{arrows}
\usetikzlibrary{decorations.pathmorphing}
\tikzset{snake it/.style={decorate, decoration=snake}}
\usetikzlibrary{shapes.misc}
\tikzset{cross/.style={cross out, draw=black, minimum size=2*(#1-\pgflinewidth), inner sep=0pt, outer sep=0pt},
cross/.default={1pt}}

\usepackage[T1]{fontenc}
\usepackage{esint}

\newcommand{\vv}[1]{\left\langle #1 \right\rangle}
\newcommand{\un}[1]{\underline{#1}}

\def\be#1\ee{\begin{align}#1\end{align}}

\begin{document}

\unitlength = .8mm

\begin{titlepage}

\begin{center}

\hfill \\
\hfill \\
\vskip 1cm

\title{Bootstrapping the Ising Model on the Lattice}

\author{Minjae Cho$^{\text{\Lightning}}$, Barak Gabai${}^{\text{\Snow}}$, Ying-Hsuan Lin${}^{\text{\Snow}}$, \\  Victor A. Rodriguez$^{\text{\FilledRainCloud}}$, Joshua Sandor${}^{\text{\Snow}}{}^{\text{\faFire}}$, Xi Yin${}^{\text{\Snow}}$}

\address{
$^{\text{\Lightning}}$Princeton Center for Theoretical Science, Princeton University, \\ Princeton, NJ 08544,
USA
\\
${}^{\text{\Snow}}$Jefferson Physical Laboratory, Harvard University, \\
Cambridge, MA 02138 USA
\\
$^{\text{\FilledRainCloud}}$Joseph Henry Laboratories, Princeton University, \\ Princeton, NJ 08544, USA 
\\
$^{\text{\faFire}}$Stanford Institute for Theoretical Physics, Stanford University, 
\\
Stanford, CA 94305, USA
}

\email{minjae@princeton.edu, bgabai@g.harvard.edu, yhlin@fas.harvard.edu, vrodriguez@princeton.edu, jsandor@fas.harvard.edu,  xiyin@fas.harvard.edu}

\end{center}

\abstract{ We study the statistical Ising model of spins on the infinite lattice using a bootstrap method that combines spin-flip identities with positivity conditions, including reflection positivity and Griffiths inequalities, to derive rigorous two-sided bounds on spin correlators through semidefinite programming. For the 2D Ising model on the square lattice, the bootstrap bounds based on correlators supported in a 13-site diamond-shaped region determine the nearest-spin correlator to within a small window, which for a wide range of coupling and magnetic field is narrower than the precision attainable with simple Monte Carlo methods. We also report preliminary results of the bootstrap bounds for the 3D Ising model on the cubic lattice.
}

\vfill

\end{titlepage}

\eject

\begingroup
\hypersetup{linkcolor=black}

\tableofcontents

\endgroup

\section{Introduction}

The Ising model, defined as a thermodynamic ensemble of spins that interact amongst nearest neighbors on a $d$-dimensional lattice $\Lambda$, with the partition function
\ie
Z = \sum_{s_x=\pm 1,~x\in\Lambda} e^{J \sum_{\langle xy\rangle} s_x s_y + h \sum_x s_x},
\fe
is a basic model of ferromagnetism. For $d=2,3$ and in the absence of external magnetic field $h$, the Ising model exhibits a second order phase transition at a critical value of the coupling (or inverse temperature) $J$, near which the long range correlations are captured by a unitary quantum field theory \cite{Wilson:1973jj}. While the $d=2$ Ising model is known to be exactly solvable at $h=0$ \cite{Onsager:1943jn, Wu:1975mw},\footnote{See \cite{McCoy:1978ta, Zamolodchikov:1989fp, Yurov:1991my, Fonseca:2001dc, Fonseca:2006au} for field theoretic investigations of the $d=2$ Ising model near criticality with nonzero external magnetic field.} the $d=3$ one is not. In recent years, the conformal bootstrap \cite{Poland:2018epd}, based on locality, positivity, and the assumption of conformal symmetry, has emerged as a powerful tool that nearly solves the $d=3$ Ising model at criticality as a conformal field theory \cite{El-Showk:2012cjh}, in the sense of rigorously determining the low lying operator spectrum and structure constants to high precision \cite{Kos:2016ysd}.

An alternative bootstrap approach, based on positivity together with certain identities that constrain expectation values and are known as loop equations \cite{Migdal:1983qrz}, has recently been applied to matrix models and lattice gauge theories at large $N$ \cite{Anderson:2016rcw, Lin:2020mme, Han:2020bkb, Kazakov:2021lel, Kazakov:2022xuh}, giving rigorous albeit numerical two-sided bounds on observables such as Wilson loop expectation values.\footnote{See \cite{goluskin2018bounding, tobasco2018optimal, goluskin2020bounding} for analogous bootstrap approach to classical dynamical systems, and \cite{han2020quantum} for that of quantum many-body systems.} Furthermore, there is evidence that as one enlarges the space of constraints, the two-sided bounds may converge and eventually pin down the exact results.

\begin{figure}[h!]
	\def\svgwidth{1\linewidth}
	\centering{
\begingroup%
  \makeatletter%
  \providecommand\color[2][]{%
    \errmessage{(Inkscape) Color is used for the text in Inkscape, but the package 'color.sty' is not loaded}%
    \renewcommand\color[2][]{}%
  }%
  \providecommand\transparent[1]{%
    \errmessage{(Inkscape) Transparency is used (non-zero) for the text in Inkscape, but the package 'transparent.sty' is not loaded}%
    \renewcommand\transparent[1]{}%
  }%
  \providecommand\rotatebox[2]{#2}%
  \newcommand*\fsize{\dimexpr\f@size pt\relax}%
  \newcommand*\lineheight[1]{\fontsize{\fsize}{#1\fsize}\selectfont}%
  \ifx\svgwidth\undefined%
    \setlength{\unitlength}{1200bp}%
    \ifx\svgscale\undefined%
      \relax%
    \else%
      \setlength{\unitlength}{\unitlength * \real{\svgscale}}%
    \fi%
  \else%
    \setlength{\unitlength}{\svgwidth}%
  \fi%
  \global\let\svgwidth\undefined%
  \global\let\svgscale\undefined%
  \makeatother%
  \begin{picture}(1,0.33333333)%
    \lineheight{1}%
    \setlength\tabcolsep{0pt}%
    \put(0,0){\includegraphics[width=\unitlength,page=1]{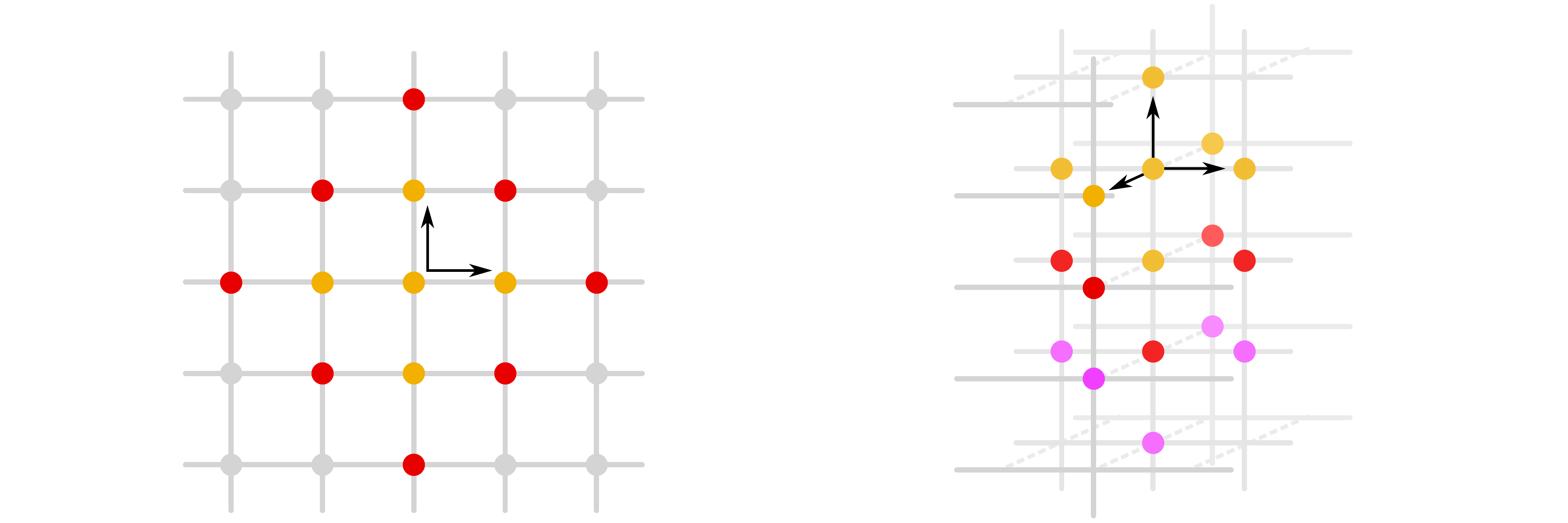}}%
    \put(0.28009918,0.18931254){\makebox(0,0)[lt]{\lineheight{1.25}\smash{\begin{tabular}[t]{l}$e_2$\end{tabular}}}}%
    \put(0.29454714,0.13385115){\makebox(0,0)[lt]{\lineheight{1.25}\smash{\begin{tabular}[t]{l}$e_1$\end{tabular}}}}%
    \put(0.74047984,0.26131694){\makebox(0,0)[lt]{\lineheight{1.25}\smash{\begin{tabular}[t]{l}$e_3$\end{tabular}}}}%
    \put(0.77352117,0.20739506){\makebox(0,0)[lt]{\lineheight{1.25}\smash{\begin{tabular}[t]{l}$e_2$\end{tabular}}}}%
    \put(0.71473668,0.19827133){\makebox(0,0)[lt]{\lineheight{1.25}\smash{\begin{tabular}[t]{l}$e_1$\end{tabular}}}}%
  \end{picture}%
\endgroup%
	\caption{Domains on the lattice. The left panel illustrates the 2D ``131'' diamond in orange within the ``13531'' diamond in red and orange. The right panel shows the 3D ``151'' diamond (orange), the ``1551'' diamond (orange and red) and the ``15551'' diamond (orange, red and pink).
			\label{domains}
	}}
\end{figure}
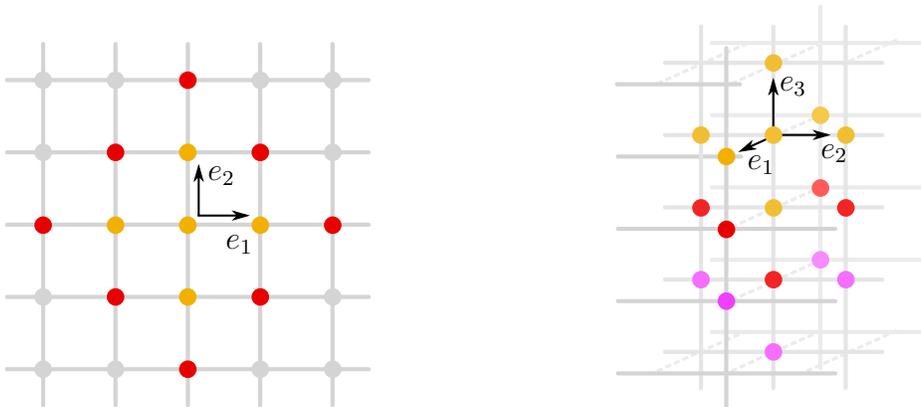

In this paper, we adopt the same strategy as that of \cite{Anderson:2016rcw, Kazakov:2022xuh} to study the Ising model on the square\footnote{The spin-flip equations and some of the positivity conditions on the square lattice considered in this paper have been independently derived and studied in unpublished work of Jiaxin Qiao and Zechuan Zheng.} and cubic lattices, focusing on spin correlation functions defined by
\ie\label{expectval}
\vv{ \un{s}_A } = {1\over Z}  \sum_{s_x=\pm 1,~x\in\Lambda} \un{s}_A\,  e^{J \sum_{\langle xy\rangle} s_x s_y+ h \sum_x s_x},~~~~~\un{s}_A \equiv \prod_{x\in A} s_x,
\fe
where $A\subset \Lambda$ is a finite set of lattice sites. We first derive the analog of loop equations, which we refer to as spin-flip equations, of the form \cite{1962PhL.....2...15D,1963PhL.....4..161C,1965PhL....19..267S,1970CMaPh..18...82B}
\ie\label{looppre}
\big( 1 - 2 \chi_A(z) \big) \vv{ \un{s}_A } = \vv{ \un{s}_A \, \exp\left[ -2J s_z \sum_{\mu=1}^d (s_{z+e_\mu} + s_{z- e_\mu}) - 2h s_z \right] },
\fe
where $e_\mu$ are unit basis vectors of the lattice, and $\chi_A$ is the characteristic function
\ie
\chi_A(z) \equiv \left\{\begin{array}{ll} 1 , &z\in A ,\\  0,~& z \not\in A. \end{array} \right.
\fe
By lattice translation invariance of expectation values, we can restrict to the case $z=0$, the origin of the lattice $\Lambda$, without loss of generality. For each subset $A$, (\ref{looppre}) can be reduced to a linear relation among a finite set of spin correlators, of the form
\ie\label{labform}
\sum_{B\subset\Lambda} L_{A|B} \vv{\un{s}_B} = 0,
\fe
where $L_{A|B}$ are real coefficients. Furthermore, given a domain ${\cal D}\subset \Lambda$, we can find a sub-domain ${\cal D}'$ such that for every $A\subset {\cal D}'$, $L_{A|B}\not=0$ only if $B\subset {\cal D}$.  For example, if $\cD$ consists of the orange and red sites depicted on the left panel of figure~\ref{domains}, then $\cD'$ consists of the yellow sites.

Next, we consider positivity constraints obeyed by spin correlators, of three types. The first type of constraints is {\it reflection positivity} \cite{Osterwalder:1977pc}, of the form
\ie\label{reflpos}
\vv{ {\cal O} \, {\cal O}^R } \geq 0,
\fe 
where 
\ie\label{odef}
{\cal O}=\sum_{A \subset H} t^A \un{s}_A,~~~~{\cal O}^R=\sum_{A \subset H} t^A \un{s}_{R(A)}.
\fe 
Here $H$ is half of the lattice defined by $H = \{x\in \Lambda: v\cdot x\geq c\}$, for a fixed primitive lattice vector $v$ and a constant $c$, and $R$ is the mirror reflection in the direction of $v$, defined by
\ie\label{rxdef}
R_{v,c}(x) = x -  {2(v\cdot x - c) \over v^2} v.
\fe
The sum in (\ref{odef}) is taken over a finite set of subsets $A$, with real coefficients $t^A$. Importantly, $v$ and $c$ must be chosen such that $R$ is an automorphism of the lattice $\Lambda$, which restricts $v^2\in \{1,2\}$ and $c\in {v^2\over 2} \mathbb{Z}$.\footnote{Given a primitive $v$, $R_{v,c}(x)$ is a lattice vector if and only if $\frac{2(v \cdot x - c)}{v^2} \in \bZ$. As $v\cdot x$ can take every integer value, $v^2$ must be either 1 or 2.} Up to lattice rotations and shifts, there are three inequivalent choices:
\ie\label{twodrefl}
&v = e_1,~~~~~~~~ c=0~~~~~~(R_{e_1, 0})
\\
& v = e_1,~~~~~~~~ c ={1\over 2}~~~~~\,(R_{e_1, {1\over 2}})
\\
& v = e_1 + e_2,~~ c= 0~~~~~\,(R_{e_1+e_2,0})
\fe
(\ref{reflpos}) is derived by rewriting the correlator in question as a sum of squares. Note that in the case of half-integer $c$, such as $R_{e_1,{1\over 2}}$ of (\ref{twodrefl}), the derivation of (\ref{reflpos}) requires the assumption of ferromagnetic coupling $J\geq 0$ (see section \ref{sec:reflect}).

\begin{figure}[h!]
	\def\svgwidth{1\linewidth}
	\centering{
\begingroup%
  \makeatletter%
  \providecommand\color[2][]{%
    \errmessage{(Inkscape) Color is used for the text in Inkscape, but the package 'color.sty' is not loaded}%
    \renewcommand\color[2][]{}%
  }%
  \providecommand\transparent[1]{%
    \errmessage{(Inkscape) Transparency is used (non-zero) for the text in Inkscape, but the package 'transparent.sty' is not loaded}%
    \renewcommand\transparent[1]{}%
  }%
  \providecommand\rotatebox[2]{#2}%
  \newcommand*\fsize{\dimexpr\f@size pt\relax}%
  \newcommand*\lineheight[1]{\fontsize{\fsize}{#1\fsize}\selectfont}%
  \ifx\svgwidth\undefined%
    \setlength{\unitlength}{1300bp}%
    \ifx\svgscale\undefined%
      \relax%
    \else%
      \setlength{\unitlength}{\unitlength * \real{\svgscale}}%
    \fi%
  \else%
    \setlength{\unitlength}{\svgwidth}%
  \fi%
  \global\let\svgwidth\undefined%
  \global\let\svgscale\undefined%
  \makeatother%
  \begin{picture}(1,0.28461538)%
    \lineheight{1}%
    \setlength\tabcolsep{0pt}%
    \put(0,0){\includegraphics[width=\unitlength,page=1]{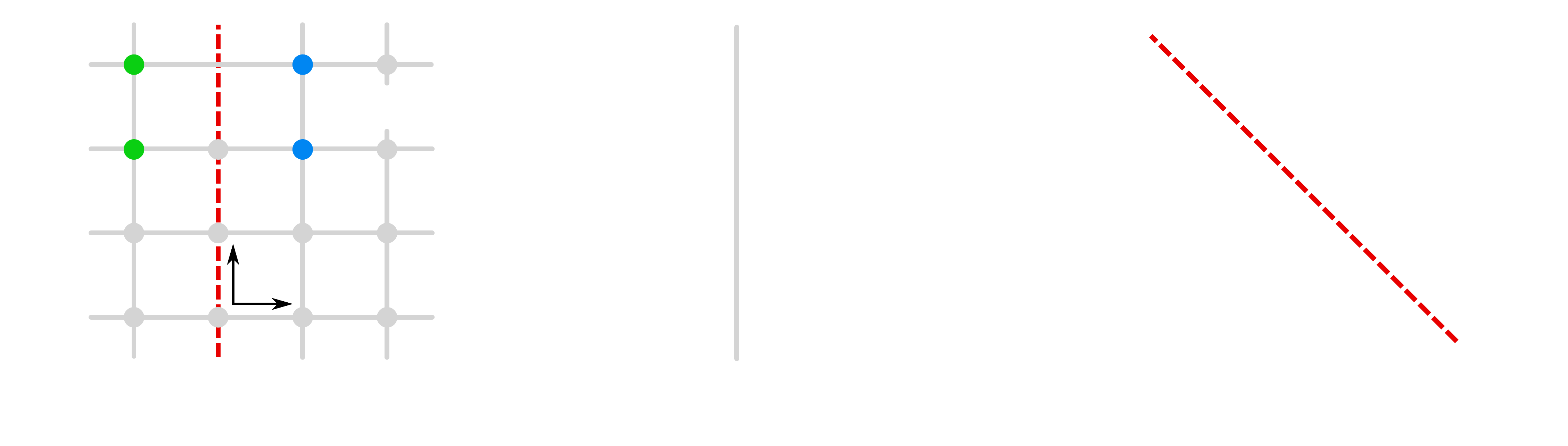}}%
    \put(0.15551201,0.10991155){\makebox(0,0)[lt]{\lineheight{1.25}\smash{\begin{tabular}[t]{l}$e_2$\end{tabular}}}}%
    \put(0.1665409,0.06448566){\makebox(0,0)[lt]{\lineheight{1.25}\smash{\begin{tabular}[t]{l}$e_1$\end{tabular}}}}%
    \put(0.05402269,0.21070851){\makebox(0,0)[lt]{\lineheight{1.25}\smash{\begin{tabular}[t]{l}$A$\end{tabular}}}}%
    \put(0,0){\includegraphics[width=\unitlength,page=2]{reflectionpos.pdf}}%
    \put(0.142737,0.01253648){\makebox(0,0)[lt]{\lineheight{1.25}\smash{\begin{tabular}[t]{l}$R_{e_1,0}$\end{tabular}}}}%
    \put(0.47134639,0.0125362){\makebox(0,0)[lt]{\lineheight{1.25}\smash{\begin{tabular}[t]{l}$R_{e_1,\frac{1}{2}}$\end{tabular}}}}%
    \put(0.79425593,0.0125362){\makebox(0,0)[lt]{\lineheight{1.25}\smash{\begin{tabular}[t]{l}$R_{e_1+e_2,0}$\end{tabular}}}}%
    \put(0.20056115,0.21070818){\makebox(0,0)[lt]{\lineheight{1.25}\smash{\begin{tabular}[t]{l}$R_{e_1,0}(A)$\end{tabular}}}}%
    \put(0,0){\includegraphics[width=\unitlength,page=3]{reflectionpos.pdf}}%
  \end{picture}%
\endgroup%
	\caption{Three different types of reflection positivity conditions on the square lattice.  }
			\label{reflectionpos}
	}
\end{figure}

The second type of constraints is {\it square positivity}, of the form
\ie\label{sqpos}
\vv{ {\cal O}^2 } \geq 0,
\fe
where 
\ie
{\cal O}= \sum_{A \subset \Lambda} t^A \un{s}_A,~~~~~ t^A\in\mathbb{R}.
\fe

The third type of constaints are {\it Griffiths inequalities}, known to hold in the ferromagnetic Ising model ($J\geq 0$) with non-negative external magnetic field ($h\geq 0$) \cite{Glimm:1987ng}.\footnote{A variety of other Ising inequalities that constrain the spin correlators are known \cite{Glimm:1987ng, Rychkov:2016mrc} but their roles will not be explored in this paper.}  They are of the form
\ie\label{firstgriff}
& \vv{ \un{s}_A } \geq 0,~~~~~~~(G_1)
\fe
and
\ie\label{secondgriff}
& \vv{\un{s}_A \un{s}_B } - \vv{\un{s}_A} \vv{ \un{s}_B } \geq 0,~~~~~~~(G_2)
\fe
for arbitrary finite subsets $A, B\subset \Lambda$. See section \ref{sec:griffine} for their derivations.

The above positivity constraints, with the exception of the second Griffiths inequality,\footnote{To recast the second Griffiths inequality (\ref{secondgriff}) as the positive-semidefiniteness of a matrix whose entries are linear in the spin correlators is only straightforward in the case where $A, B$ are related by symmetries of the lattice.} can be formulated as the positive-semidefiniteness of a matrix
\ie
\left( {\cal M}_{IJ}\right)  \succeq 0,
\fe
whose entries are linear in the spin correlators $\vv{\un{s}_A}$, of the form
\ie
{\cal M}_{IJ} = \sum_{A\subset \Lambda} M_{IJ|A} \vv{\un{s}_A}.
\fe
Given a domain ${\cal D}\subset\Lambda$, we can restrict the indices $I,J$ to be such that $M_{IJ|A}$ is nonzero only for $A\subset {\cal D}$; the resulting restricted matrix $({\cal M}_{IJ})$ will be denoted ${\cal M}^{(D)}$, and the corresponding coefficients $M_{IJ|A}^{(\cD)}$. 

We can now obtain a lower bound on a given spin correlator $\vv{\cal O} = \sum_A t^A \vv{\un{s}_A}$ by solving the semidefinite programming (SDP) problem of finding
\ie\label{sdpprob}
& \min_{x_A\in\mathbb{R},~A\subset {\cal D}} \sum_A t^A x_A
\\
&{\rm subject~to~} ~x_\emptyset=1, ~~ \sum_{B\subset {\cal D}} L_{A|B} x_B = 0, ~~(A\subset {\cal D}')
\\
&~~~~~~~~{\rm and}~ \left(  \sum_{A\subset {\cal D}} M^{(\cD)}_{IJ|A} x_A \right) \succeq 0.
\fe
Similarly, an upper bound on $\vv{O}$ is obtained by replacing $t^A$ with $-t^A$ in (\ref{sdpprob}), and flipping the sign of the minimization result. Section \ref{sec:invsdp} describes a key technical improvement on the computational efficiency that utilizes the representation theory of the lattice automorphism group and the technology of \emph{invariant SDP} \cite{bachoc2012invariant}.

We have implemented reflection positivity (1.11) and the first Griffiths inequality (1.13) through SDP to obtain two-sided bounds. Square positivity, on the other hand, appears to be redundant.\footnote{Assuming that the domain $\cD$ is chosen appropriately so that all of its sites appear in the spin-flip equations on the domain.} The role of the second Griffiths inequality, which cannot be straightforwardly implemented through SDP, is discussed in section \ref{sec:secondgriff}. For the 2D Ising model on the square lattice, with the region $\cD$ taken to be the 13531 diamond (Figure \ref{domains}), the bootstrap bound is seen to pin down the correlator of nearest spins $\vv{s_0 s_{e_1}}$ (or energy density) to within a window whose width is highly sensitive to the value of the coupling. See Figure \ref{fig:h0S11} for a comparison of the bootstrap bounds versus the exact answer in the $h=0$ case, and Figure \ref{fig:hS11} for the analogous bounds at nonzero $h$. 

Compared to the estimated error bars of the Monte Carlo simulation, the bootstrap bounds we have obtained thus far are wider in the vicinity of the critical coupling, but quickly narrow away from criticality and can be smaller than the precision attainable with simple Monte Carlo methods (see Figure \ref{fig:logplotS11} and \ref{fig:hS11err}).  We emphasize that unlike Monte Carlo methods which are subject to errors due to statistical uncertainty and finite size effects, our bootstrap bounds on the spin correlators are rigorous results for the Ising model on the \emph{infinite} lattice, where the numerical precision of the SDP solver only affects the extent to which the upper and lower bounds are optimized.

For the 3D Ising model on the cubic lattice, we have worked with $\cD$ up to the 1551 domain (Figure \ref{domains}), as well as the 15551 domain with a truncated (nonetheless rigorous) version of the reflection positivity conditions.\footnote{The truncation is adopted due to a computational bottleneck of the SDP solver used in this work, which we hope to overcome in the near future.} The resulting bootstrap bound, as shown in Figure \ref{fig:3dS11bounds}, is seen to be consistent with the results from Monte Carlo simulations.

In section \ref{sec:loop}, we will express the spin-flip identities as linear constraints among correlators, and describe the algorithm for solving these constraints. The positivity conditions are derived in section \ref{sec:allposcond}. In section \ref{sec:invsdp}, we formulate the positivity conditions as an SDP problem. Details of the bootstrap bounds on spin correlators are presented in section \ref{sec:allbounds}.

A key question concerns the convergence of the bootstrap bounds on the spin correlators, and in particular whether the rate of convergence near the critical coupling is sufficiently fast to be useful for extracting quantum field theoretic observables in a way that is complimentary to the conformal bootstrap. This is closely tied to the efficiency of the algorithms in solving the spin-flip identities and implementing the positivity constraints, as well as in identifying the most relevant sets of spin configurations on domain. These issues will be discussed in the concluding section \ref{sec:discuss}.

\section{Spin-flip equations}
\label{sec:loop}


We start with the identity (\ref{looppre}), which follows from redefining the variable $s_z \to -s_z$ at a single site $z$ in evaluating the sum on the RHS of (\ref{expectval}). As already stated, it suffices to restrict to the case of $z=0$. To reduce (\ref{looppre}) to an equation that involves a finite set of correlators, let us consider the variable
\ie
w &\equiv \sum_{\mu=1}^d (s_{e_\mu} + s_{-e_\mu})  \in \{0,\pm 2,\cdots, \pm 2d\},
\fe
which obeys the degree $2d+1$ polynomial equation
\ie\label{pweq}
P(w) \equiv \prod_{k=-d}^d (w-2k) = 0.
\fe
It follows that we can replace all $w^n$ for $n\geq 2d+1$ with lower degree polynomials in $w$, of the form
\ie\label{wzrel}
w^{n} = \sum_{\ell=0}^{2d} C_{n,\ell}w^\ell,
\fe
where $C_{n,\ell}=\delta_{n\ell}$ for $n\leq 2d$. We can rewrite  (\ref{looppre}) as
\ie
 & \vv{\un{s}_A \left[ 2 \chi_A(0)  -1 + e^{-2h s_0} \right] } 
 \\ & + \sum_{k=1}^{\infty} {(2J)^{2k}\over (2k)!} \vv{\un{s}_A w^{2k} e^{-2hs_0} } - \sum_{k=0}^\infty {(2J)^{2k+1}\over (2k+1)!} \vv{ \un{s}_A s_0 w^{2k+1} e^{-2hs_0} } = 0,
\fe
or equivalently using (\ref{wzrel}),
\ie\label{loopeqnab}
& \left[ 2 \chi_A(0) - 1 + \cosh(2h)\right] \vv{\un{s}_A } + \sum_{\ell=0}^{2d} \left[ A_\ell \cosh(2h) + B_\ell \sinh(2h) \right] \vv{\un{s}_A w^{\ell} } 
\\
&  - \sinh(2h) \vv{\un{s}_A s_0}
- \sum_{\ell=0}^{2d} \left[ A_\ell \sinh(2h) + B_\ell \cosh(2h) \right] \vv{\un{s}_A s_0 w^{\ell} } = 0,
\fe
where $A_\ell$ and $B_\ell$ are a set of constants that depend on the lattice dimension $d$ and the coupling $J$, given by
\ie\label{abforms}
A_\ell \equiv \sum_{k=1}^{\infty} {(2J)^{2k}\over (2k)!}  C_{2k,\ell},
~~~~~ B_\ell \equiv  \sum_{k=0}^\infty {(2J)^{2k+1}\over (2k+1)!}  C_{2k+1,\ell} .
\fe
Explicit formulae for these coefficients are given in Appendix \ref{sec:loopcoeff}. In particular, the non-vanishing $A_\ell$ and $B_\ell$ coefficients appearing in the spin-flip equation (\ref{loopeqnab}) are $A_2, A_4, \cdots, A_{2d}$ and $B_1, B_3,\cdots, B_{2d-1}$.

\subsection{A 2D example}
\label{sec:2dexloop}

Consider the $d=2$ Ising model with $h=0$. The spin-flip equation (\ref{loopeqnab}) for $A=\emptyset$ is explicitly
\ie\label{firstloopeqn}
0 &= \sum_{\ell=0}^{4}  A_\ell  \vv{ w^{\ell} } - \sum_{\ell=0}^{4}  B_\ell \vv{ s_0 w^{\ell} } 
\\
&= A_2\left(4+4\vv{s_{e_1} s_{-e_1}} + 8 \vv{s_{e_1} s_{e_2}}\right) 
\\
&~~~ + A_4 \left( 40 + 64 \vv{s_{e_1} s_{-e_1}}
+ 128 \vv{s_{e_1} s_{e_2}}  +24 \vv{s_{e_1} s_{-e_1} s_{e_2} s_{-e_2}} \right)
\\
&~~~ - 4 B_1 \vv{s_0 s_{e_1}} - B_3 \left( 40  \vv{s_0 s_{e_1}} + 24 \vv{s_0 s_{e_1} s_{-e_1} s_{e_2}} \right) ,
\fe
where the coefficients $A_2, A_4, B_1, B_3$ are functions of $J$ given in (\ref{abdtwocase}). In writing the second equality, we have used the symmetries of the lattice as well as $\vv{1}=1$. The spin correlators involved in (\ref{firstloopeqn}) are
\ie\label{xcorrs}
& x_1 = \vv{s_0 s_{e_1}},~~~ x_2 =  \vv{s_{e_1} s_{-e_1}},~~~ x_3 = \vv{s_{e_1} s_{e_2}},
\\
& x_4 = \vv{s_{e_1} s_{-e_1} s_{e_2} s_{-e_2}},~~~ x_5 = \vv{s_0 s_{e_1} s_{-e_1} s_{e_2}}.
\fe
There are 5 more spin-flip equations that involve the same set of correlators $x_1,\cdots, x_5$, corresponding to $A=\{0,e_1\}$, $\{\pm e_1\}$, $\{e_1, e_2\}$, $\{\pm e_1, \pm e_2\}$, $\{0, \pm e_1, e_2\}$. Without taking into account the specific values of $A_2, A_4, B_1, B_3$, one finds that two of these 6 spin-flip equations are not independent and can be discarded. Remarkably, the \emph{nonlinear} relations satisfied by $A_\ell,B_\ell$ in (\ref{abdtwocase}) are such that only 2 out of the remaining 4 spin-flip equations are independent. From these equations we can determined two of the correlators in (\ref{xcorrs}) in terms of the other three,
\ie{}
& x_4 = { - 8 (\cosh(2 J) + \cosh(6 J)) x_1 + \sinh(2 J) ( -1 + 2 x_2 + 4 x_3) +
\sinh(6 J) (3 + 2 x_2 + 4 x_3)\over 4 \sinh^3(2J)},
\\
& x_5 = {-(1 + 3 \cosh(4 J)) x_1 + \sinh(4 J) (1 + x_2 + 2 x_3) \over 2\sinh^2(2J)}.
\fe

\subsection{A 3D example} 

Consider the $d=3$ Ising model with $h=0$ on the 151 diamond $\cD=\{0,\pm e_1, \pm e_2, \pm e_3\}$ (the orange sites on the right panel of Figure~\ref{domains}). In this case, the independent spin correlators are (\ref{xcorrs}) together with
\ie\label{xcorrs3d}
& x_6 = \vv{s_{e_1} s_{-e_1} s_{e_2} s_{e_3}},~~~ x_7 =  \vv{s_0 s_{e_1} s_{e_2} s_{e_3}},
\\
&  x_8 = \vv{s_{e_1}s_{-e_1} s_{e_2}s_{-e_2} s_{e_3}s_{-e_3}},~~~ x_9 = \vv{ s_0 s_{e_1} s_{-e_1} s_{e_2} s_{-e_2} s_{e_3}  }.
\fe
There are 10 spin-flip equations involving the set of correlators $x_1,\cdots, x_9$, corresponding to to $A = \emptyset, \{0,e_1\}$, $\{\pm e_1\}$, $\{e_1, e_2\}$, $\{\pm e_1, \pm e_2\}$, $\{0, \pm e_1, e_2\}$, $\{\pm e_1, e_2,e_3\}$, $\{ 0,e_1,e_2,e_3 \}$, $\{ \pm e_1, \pm e_2, \pm e_3 \}$, $\{ 0, \pm e_1, \pm e_2, e_3 \}$. For instance, the spin-flip equation for the case $A = \emptyset$ is illustrated as 
\ie
\includegraphics[width=0.91\textwidth]{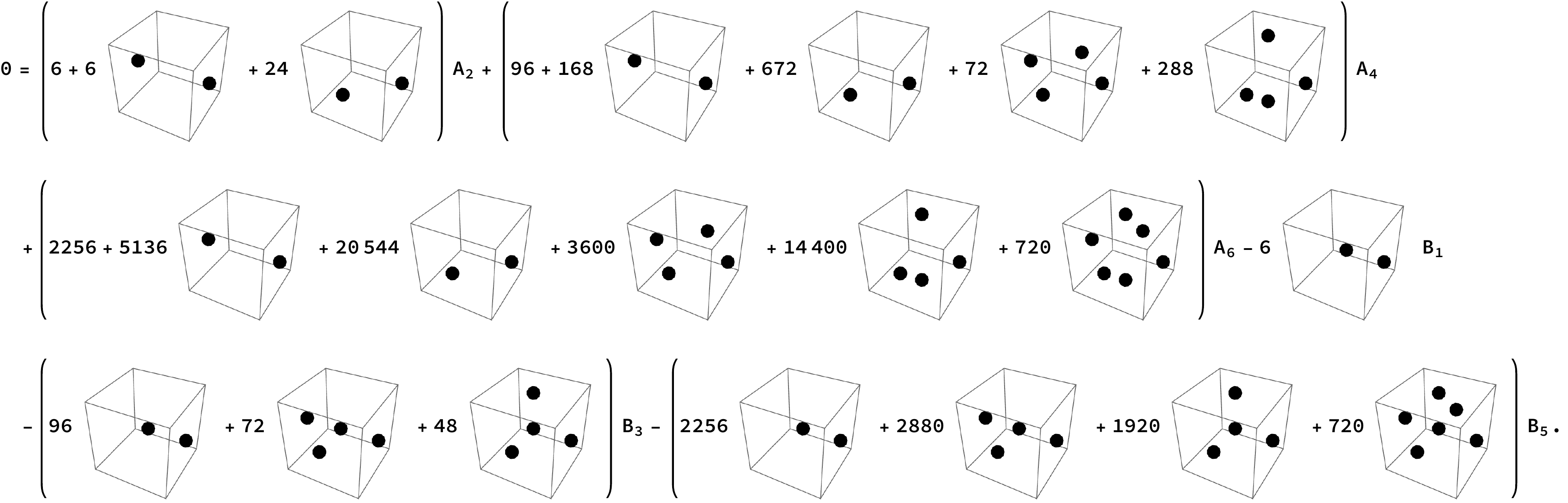}
\label{eq:3dloop0sample}
\fe
After taking into account the explicit expressions for the coefficients $A_l$ and $B_l$ for $d=3$ in Appendix \ref{sec:loopcoeff}, it turns out that only 4 out of the 10 spin-flip equations are independent. From these, we can determine the spin correlators 
\ie
x_5 &= -\frac{\cosh^3(2 J) \sinh^2(2 J) }{\sinh^4(4J)} \Big[ 4 x_1 \cosh(2 J) + 4 x_1 \cosh(6 J)  
\\
&~~~~ - 2 (1 + 4 x_3 - x_4 + (1 + 2 x_2 + 4 x_3 + x_4) \cosh(4 J)) \sinh(2 J)  \Big], 
\\
x_7 &= -\frac{4 \cosh^5(2 J) \sinh^4(2 J) }{\sinh^6(4 J)} \Big[ 4 x_1 \cosh(2 J) + 4 x_1 \cosh(6 J)  
 \\
& ~~~~ - 2 (1 + 2 x_2 + 2 x_3 + x_4 - 2 x_6 + (1 + 6 x_3 - x_4 + 2 x_6) \cosh(4 J)) \sinh(2 J) \Big], 
\\
x_8 &= -\frac{1}{8\sinh^5(2J)} \Big[ -8 x_1 (\cosh(2 J) + \cosh(10 J)) + (3 + x_2 + 4 x_3 + x_4 + 4 x_6)\sinh(2 J)  
\\
&~~~~ + ( 4 (x_2 + 4 x_3) \cosh(4 J) + (5 + 3 x_2 + 12 x_3 - x_4 - 4 x_6) \cosh(8 J)) \sinh(2 J) \Big], 
\\
x_9 &= \frac{1}{16 \sinh^4(2J)}\Big[ 6 x_1 + 10 x_1 \cosh(8 J) - 2 (-1 + x_4 + 4 x_6) \sinh(4 J) 
\\
&~~~~ + (-3 - 2 x_2 - 8 x_3 + x_4 + 4 x_6) \sinh(8 J) \Big].
\label{eq:loopeqs3d151sol}
\fe
Note that on the 151 diamond the spin-flip equations determine almost all intrinsically $d=3$ spin correlators (\ref{xcorrs3d}), except for one ($x_6$ in the case shown in (\ref{eq:loopeqs3d151sol})) which remains a free variable.

\subsection{Algorithm for generating and solving spin-flip equations}
\label{sec:algloop}

For larger regions $\cD$, we start by enumerating all primitive subsets $A\subset\cD$ unrelated by symmetries of the lattice, and generate the spin-flip equations from the analog of (\ref{loopeqnab}) with the origin (location of spin flip) shifted to points $x\in \cD$ that obey
\ie\label{xxind}
x, x\pm e_\mu \in \cD,~~~~\mu=1,\cdots,d.
\fe 
For the 2D 13531 diamond region $\cD$ (the orange and red sites on the left panel of Figure~\ref{domains}), in the case $h=0$ with the assumption of unbroken spin-flip $\mathbb{Z}_2$ symmetry, there are 2656 such spin-flip equations associated with the 569 primitive subsets of $\cD$, out of which 549 equations turn out to be independent, determining all but 19 spin correlators $\vv{\un{s}_A}$ with $A\subset\cD$. In this case, an analytic solution to the spin-flip equations as a function of $J$ is practically unattainable. Instead, we solve the equations numerically at each given value of $J$. Note that sufficiently high numerical precision is required to identify the correct number of linearly independent spin-flip equations.

The numbers of primitive subsets (including the empty set), independent spin-flip equations, and independent spin correlators after solving the spin-flip equations, for the 2D 13531 diamond at nonzero $h$ without $\mathbb{Z}_2$ symmetry, and for the 3D 1551 and 15551 regions at $h=0$ with the assumption of unbroken $\mathbb{Z}_2$ symmetry, are summarized in the following table.

\begin{center}
        \begin{tabular}{ |c | c | c | c|  }
        \hline
         & primitive subsets   & ind. spin-flip equations& ind. spin correlators  \\
                    \hline
 2D 13531, $h=0$   & 569   &  549  & 19 \\
                    \hline
 2D 13531, $h\not=0$  & 1127 & 1097 & 29 \\
                     \hline
 3D 1551, \, $h=0$  & 214 & 162  & 51 \\
                      \hline
 3D 15551, $h=0$  & 5214 & 4584  & 629 \\
        \hline
        \end{tabular}
 \end{center}

Given $m$ linear spin-flip equations for $n$ spin correlators (including $\vev{\un{s}_\emptyset} = 1$), we can package them as a matrix equation $L s = 0$, where $s$ is an $n$-dimensional vector and $L$ is an $m\times n$ matrix of rank $r$.\footnote{For instance, in the 2D 13531 $h = 0$ case, $m = 2656$, $n = 569$, and $r = 549$.}  The solution space to the spin-flip equations is the kernel of $L$.  We choose $n-r$ out of the $n$ spin correlators to be a basis for the kernel, and solve the remaining $r$ spin correlators in this basis. As $L$ is a (highly) degenerate numerical matrix, naive linear algebraic algorithms such as row reduction 
and Gram-Schmidt 
can suffer numerical instability when the system size is large.  To circumvent this, we first perform a QR decomposition (resulting in $L = QR$, with $Q$ orthogonal and $L$ upper-triangular) via numerically stable algorithms such as Householder reflections with pivoting, and then transform $R$ to row echelon form.\footnote{Gram-Schmidt theoretically achieves QR decomposition, but is practically unstable.  To exemplify the importance of algorithm, let us mention that in the 3D 15551 $h = 0$ case, solving the spin-flip equations with the Mathematica Solve command typically requires 50-digit working precision and takes about half a day, but with numerically stable QR algorithms available in e.g.\,Julia's LinearAlgebra package, the computation only requires double precision ($\sim 16$ digits) and takes 90 seconds.}

\section{Positivity conditions}
\label{sec:allposcond}

\subsection{Reflection positivity}
\label{sec:reflect}

To prove reflection positivity (\ref{reflpos}), let us first consider the case $c=0$, where $H$ and $R(H)$ meet along the co-rank 1 lattice $P=\{x\in\Lambda: v\cdot x=0\}$. This includes the horizontal reflection $(R_{e_1, 0})$ and the diagonal reflection $(R_{e_1+e_2,0})$ of (\ref{twodrefl}). The correlator in question can be written as
\ie\label{oopp}
\vv{ {\cal O} \, {\cal O}^R } =  \sum_{s_x=\pm1,~x\in P} e^{-J \sum_{\langle xy\rangle \subset P} s_x s_y - h \sum_{x\in P} s_x} \big( \Psi[ \un{s}_P ] \big)^2,
\fe
where
\ie\label{psipf}
\Psi[\un{s}_P] = {1\over \sqrt{Z}} \sum_{s_x=\pm1, ~ x\in H\backslash P} {\cal O}\, \exp\left[ J \sum_{\langle xy\rangle \subset H} s_x s_y + h \sum_{x\in H} s_x \right].
\fe
Note that replacing $H$ by $R(H)$ on the RHS of (\ref{psipf}) produces an identical result. The key in deriving (\ref{oopp}) is the fact that every link $\langle xy\rangle$ is either contained in $H$ or in $R(H)$. (\ref{oopp}) is a sum of non-negative terms, and is thus obviously non-negative.


The remaining case $v^2=1$, $c={1\over 2}$, e.g. $(R_{e_1, {1\over 2}})$ of (\ref{twodrefl}), requires a different treatment. In this case, $H\cap R(H) = \emptyset$, and furthermore there are links that are contained in neither $H$ nor $R(H)$. Let us define $P_+ = \{x\in\Lambda: v\cdot x = 1\}$ and $P_- = \{x\in\Lambda: v\cdot x = 0\}$, which lie on the boundary of $H$ and $R(H)$ respectively.
We can write
\ie\label{ooppq}
\vv{ {\cal O} \, {\cal O}^R } = \sum_{s_x=\pm1,~x\in P_+\cup P_-} \exp\left[ J \sum_{\langle xy\rangle,\, x\in P_+,\,y\in P_-} s_x s_y  \right]  \Psi[ \un{s}_{P_+} ] \Psi[ \un{s}_{P_-} ] ,
\fe
where
\ie
\Psi[ \un{s}_{P_+} ] = {1\over \sqrt{Z}}  \sum_{s_x=\pm1, ~ x\in H\backslash P_+} {\cal O} \exp\left[ J \sum_{\langle xy\rangle \subset H} s_x s_y + h \sum_{x\in H} s_x \right],
\fe
and $\Psi[ \un{s}_{P_-} ]$ is given by the identical function with the argument $\un{s}_{P_+}$ replaced by $\un{s}_{P_-}$. We can further write (\ref{ooppq}) equivalently as
\ie\label{oorjpos}
\vv{ {\cal O} \, {\cal O}^R } = \sum_{k=0}^\infty {J^k\over k!} \sum_{ x_1,\cdots,x_k\in P_+} \left( \sum_{s_x=\pm1,\, x\in P_+} \prod_{i=1}^k s_{x_i} \Psi[ \un{s}_{P_+} ] \right)^2 ,
\fe
which is clearly non-negative provided $J\geq 0$.

\subsection{Square positivity}

The square positivity property (\ref{sqpos}) obviously holds. Unlike reflection positivity, however, the square positivity does not have a straightforward analog in Euclidean axioms of quantum field theories based on local operators \cite{Osterwalder:1973dx, Osterwalder:1974tc, Glimm:1987ng}.

In simple examples such as the 131 diamond, where the domain $\cD$ is chosen such that every site $y\in\cD$ appears in a spin-flip identity on $\cD$, i.e. $y$ is adjacent to a site $x$ that obeys (\ref{xxind}), we have found the square positivity condition to be redundant after imposing reflection positivity.

\subsection{Griffiths inequalities}
\label{sec:griffine}

Let us recap the derivation of the Griffiths inequalities (\ref{firstgriff}) and (\ref{secondgriff}). The LHS of (\ref{firstgriff}), namely (\ref{expectval}), can be written equivalently as
\ie\label{griffsfit}
\vv{\un{s}_A} = \lim_{L\to \infty} {Z^{(L)}_{J=h=0}\over Z^{(L)}} \vv{ \un{s}_A e^{J \sum_{\langle xy\rangle} s_x s_y+ h \sum_x s_x} }^{(L)}_{J=h=0},
\fe
where $Z^{(L)}$ is the partition function of Ising model defined on the lattice $\Lambda$ with cutoff at size $L$ (say with Dirichlet boundary condition), and similarly $\vv{\cdots}^{(L)}$ the expectation value in the Ising model with IR cutoff $L$. The subscript $(J=h=0)$ refers to the Ising model with $J$ and $h$ set to zero. The expectation value appearing on the RHS of (\ref{griffsfit}) can be expanded in $J$ and $h$ as a sum over terms proportional to $\vv{ \un{s}_B }_{J=h=0}$ with positive coefficients, for some subsets $B$ of the cut-off lattice. The latter vanishes unless $B=\emptyset$, in which case $\vv{1}_{J=h=0}=1$. This proves that (\ref{griffsfit}) is non-negative.

To see (\ref{secondgriff}), we first apply a similar argument to two decoupled copies of Ising model, whose spin variables are denoted $s_x$ and $s'_x$ respectively. Define $s^\pm_x \equiv s_x\pm s'_x$, and $\un{s}^\pm_A\equiv \prod_{x\in A} s_x^\pm$. We have
\ie\label{ssvvapre}
\vv{ \un{s}^+_A \un{s}^-_B } = \lim_{L\to \infty} \left[ {Z^{(L)}_{J=h=0}\over Z^{(L)}} \right]^2 \vv{ \un{s}^+_A \un{s}^-_B e^{ {J\over 2} \sum_{\langle xy\rangle} \left( s_x^+ s_y^+ + s_x^- s_y^- \right) + h \sum_x s_x^+} }^{(L)}_{0},
\fe
where $\vv{\cdots}_0^{(L)}$ stands for expectation value in two copies of Ising model with $J=h=0$ and IR cutoff $L$. Evidently, due to the symmetry of exchanging the two copies of the model, as well as flipping sign of the second copy at $h=0$,
\ie
\vv{ \un{s}^+_A \un{s}^-_B }_0^{(L)} = \vv{ \un{s}^+_A (-)^{|B|} \un{s}^-_B }_0^{(L)} = (-)^{|A|+|B|} \vv{ \un{s}^+_A \un{s}^-_B }_0^{(L)}
\fe
is nonzero only if both $|A|$ and $|B|$ are even, and is positive in the latter case. It follows that the expansion of (\ref{ssvvapre}) in $J$ and $h$ is a sum of non-negative quantities, and thus
\ie\label{ssbasicjr}
\vv{ \un{s}^+_A \un{s}^-_B } \geq 0.
\fe
We can then write the LHS of (\ref{secondgriff}) as
\ie\label{sfsongi}
\vv{\un{s}_A \un{s}_B } - \vv{\un{s}_A} \vv{ \un{s}_B } &= \vv{\un{s}_A (\un{s}_B-\un{s}'_B)}
\\
&= \vv{ \prod_{x\in A} {s_x^+ + s_x^-\over 2}  \left( \prod_{y\in B} {s_y^+ + s_y^-\over 2} - \prod_{y\in B} {s_y^+ - s_y^-\over 2} \right)} .
\fe
The expectation value in the second line can be expanded as positive linear combinations of terms of the form $\vv{ \un{s}_{A'}^+ \un{s}_{B'}^-}$ for some $A', B'$, each of which is non-negative by (\ref{ssbasicjr}), hence (\ref{secondgriff}) follows.


\section{Invariant semidefinite programming}
\label{sec:invsdp}

A major improvement in computational efficiency is achieved by exploiting the automorphism $G$ of the lattice $\Lambda$ or a domain of the lattice, through ``invariant semidefinite programming'' \cite{bachoc2012invariant}, as follows. Suppose the domain $\cD$ (in the case of square positvity) or $\cD\cap H$ (in the case of reflection positivity) is $G$-invariant. Denote by $A^g$ the subset of $\Lambda$ obtained by acting on $A$ with an element $g\in G$. The construction of ${\cal M}_{IJ}$ is such that
\ie
M_{IJ|A^g} = \sum_{I',J'} \rho^R_{II'}(g) \rho^{R^*}_{JJ'}(g) M_{I'J'|A},
\fe
where $\rho^R(g)$ is the matrix of $g$ in some representation $R$ of $G$. Writing $R=\bigoplus_{k=1}^N R_k^{\oplus n_k}$ for distinct irreducible representations $R_i$ and multiplicity $n_i$, we can work with a basis in which $\cM$ takes the block diagonal form $\cM=\bigoplus_{k=1}^N \cM^{(k)}$, such that the matrix elements of $\cM^{(k)}= \left(\sum_A M^{(k)}_{ij,IJ|A} \vv{\un{s}_A}\right)_{1\leq i,j\leq n_k,\,1\leq I,J\leq \dim R_k}$ transform under the group action as
\ie{}
& M^{(k)}_{ij,IJ|A^g} = \sum_{I',J'=1}^{\dim R_k} \rho^{R_k}_{II'}(g) \rho^{R^*_k}_{JJ'}(g) M^{(k)}_{ij,I' J'|A}.
\fe
Since $\vv{\un{s}_{A^g}} = \vv{\un{s}_A}$, it follows from Schur's lemma that $X^{(k)}_{ij,IJ} \equiv \sum_{A\subset\cD} M^{(k)}_{ij,IJ|A}  \vv{\un{s}_A}$ is of the form $X_{ij}^{(k)} \delta_{IJ}$, and that the semidefiniteness condition in (\ref{sdpprob}) is equivalent to
\ie
\left( X^{(k)}_{ij} \right)_{1\leq i,j\leq n_k} \succeq 0,~~~k=1,\cdots,N.
\fe

\subsection{A 2D example}

Let us consider the $d=2$, $h=0$ Ising model, and restrict to correlators with spin insertions in the 131 diamond $\cD=\{0,\pm e_1, \pm e_2\}$. The independent correlators are given in (\ref{xcorrs}). On the domain $\cD$, the reflection positivity conditions $(R_{e_1, {1\over 2}})$ and $(R_{e_1+e_2, 1})$ give
\ie\label{p2q2cond}
& x_1\geq 0 ~~~~~~~~~~~~~ (R_{e_1, {1\over 2}})
\\
& \begin{pmatrix} 1 & x_3 \\ x_3 & 1 \end{pmatrix} \succeq 0~~~~(R_{e_1+e_2, 1})
\fe
Note that with spins restricted to this domain, the condition $(R_{e_1+e_2, 1})$ is independent from that of the diagonal reflection based at the origin, $(R_{e_1+e_2, 0})$, to be discussed below.

\begin{figure}[h!]
	\def\svgwidth{1\linewidth}
	\centering{
\begingroup%
  \makeatletter%
  \providecommand\color[2][]{%
    \errmessage{(Inkscape) Color is used for the text in Inkscape, but the package 'color.sty' is not loaded}%
    \renewcommand\color[2][]{}%
  }%
  \providecommand\transparent[1]{%
    \errmessage{(Inkscape) Transparency is used (non-zero) for the text in Inkscape, but the package 'transparent.sty' is not loaded}%
    \renewcommand\transparent[1]{}%
  }%
  \providecommand\rotatebox[2]{#2}%
  \newcommand*\fsize{\dimexpr\f@size pt\relax}%
  \newcommand*\lineheight[1]{\fontsize{\fsize}{#1\fsize}\selectfont}%
  \ifx\svgwidth\undefined%
    \setlength{\unitlength}{1300bp}%
    \ifx\svgscale\undefined%
      \relax%
    \else%
      \setlength{\unitlength}{\unitlength * \real{\svgscale}}%
    \fi%
  \else%
    \setlength{\unitlength}{\svgwidth}%
  \fi%
  \global\let\svgwidth\undefined%
  \global\let\svgscale\undefined%
  \makeatother%
  \begin{picture}(1,0.22307692)%
    \lineheight{1}%
    \setlength\tabcolsep{0pt}%
    \put(0,0){\includegraphics[width=\unitlength,page=1]{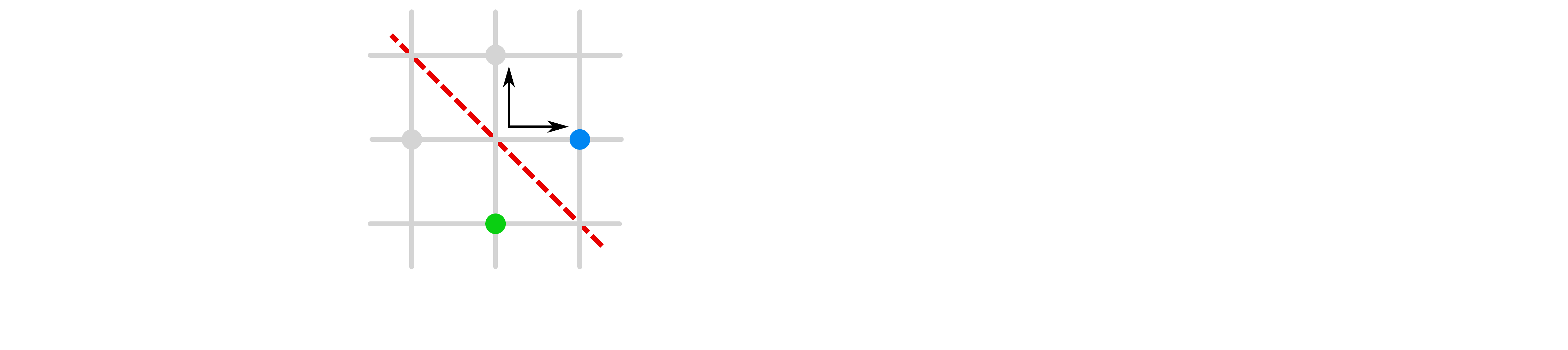}}%
    \put(0.27570594,0.01215158){\makebox(0,0)[lt]{\lineheight{1.25}\smash{\begin{tabular}[t]{l}$R_{e_1+e_2,0}$\end{tabular}}}}%
    \put(0,0){\includegraphics[width=\unitlength,page=2]{reflectiondiag.pdf}}%
    \put(0.6243835,0.01215175){\makebox(0,0)[lt]{\lineheight{1.25}\smash{\begin{tabular}[t]{l}$R_{e_1+e_2,1}$\end{tabular}}}}%
    \put(0,0){\includegraphics[width=\unitlength,page=3]{reflectiondiag.pdf}}%
  \end{picture}%
\endgroup%
	\caption{Two inequivalent diagonal reflections on the 131 diamond.
			\label{diagnonalref}
	}}
\end{figure}

To implement the reflection positivity condition $(R_{e_1,0})$, it suffices to consider linear combinations of spin insertions $\un{s}_A$ on the right side of $\cD$ with $A\subset \{0, e_1, \pm e_2\}$,  for $|A|=0,2,4$, that are even and odd with respect to the vertical reflection respectively:
\ie
& {\rm even}: ~ 1, ~s_0 s_{e_1},~s_{e_2} s_{-e_2},~  {s_0 s_{e_2} + s_0 s_{-e_2}\over 2}, ~  {s_{e_1} s_{e_2} + s_{e_1} s_{-e_2}\over 2},
~ s_0 s_{e_1} s_{e_2} s_{-e_2};
\\
& {\rm odd}:~{s_0 s_{e_2} - s_0 s_{-e_2}\over 2},~  {s_{e_1} s_{e_2} - s_{e_1} s_{-e_2}\over 2}.
\fe
The positivity of the correlator of the even and odd spin insertions with their horizontal mirror reflection gives the conditions
\ie{}
& \begin{pmatrix} 1 & x_1 & x_2 & x_1 & x_3 & x_5 \\ x_1 & x_2 & x_5 & x_3 & x_5 & x_4
\\ x_2 & x_5 & 1 & x_1 & x_3 & x_1 \\ x_1 & x_3 & x_1 & {1+x_2\over 2} & {x_1+x_5\over 2} &  x_3
\\ x_3 & x_5 & x_3 & {x_1+x_5\over 2} & {x_2 + x_4\over 2} & x_5  \\ x_5 & x_4 & x_1 & x_3 & x_5 & x_2  \end{pmatrix} \succeq 0 ~~~{\rm and}~~ \begin{pmatrix} {1-x_2\over 2}  & {x_1 - x_5\over 2}  \\ {x_1-x_5\over 2} & {x_2 - x_4\over 2}  \end{pmatrix} \succeq 0~~~~~(R_{e_1,0})
\fe
Similarly, to implement the diagonal reflection positivity $(R_{e_1+e_2,0})$, we consider the linear combinations of spin insertions $\un{s}_A$ on the upper-right side of $\cD$ with $A \subset \{0, e_1, e_2\}$, for $|A|=0,2$, that are even and odd with respect to the diagonal reflection along the 45-degree line:
\ie{}
& {\rm even}:~ 1,~s_{e_1} s_{e_2},~{s_0s_{e_1} + s_0s_{e_2}\over 2};
\\
& {\rm odd:}~{s_0s_{e_1} - s_0s_{e_2}\over 2}.
\fe
The positivity of the correlators of these spin insertions and their diagonal mirror reflection (along the 135-degree line) gives
\ie{}
& \begin{pmatrix} 1 & x_3 & x_1 \\ x_3 & x_4 & x_5 \\ x_1 & x_5 & {x_2+x_3\over 2} \end{pmatrix}\succeq 0~~~{\rm and}~~ {x_3-x_2\over 2} \geq 0~~~~~~(R_{e_1+e_2,0})
\fe

To implement square positivity, we first classify $\un{s}_A$ for $A\subset\cD$ with $|A|=0,2,4$ according to representations with respect to the dihedral automorphism group
\ie
G = \vv{ a, x \big| a^4=x^2=1,~ xax^{-1} = a^{-1} },
\fe
where $a$ is rotation by 90 degrees, and $x$ is the horizontal reflection that takes $(e_1,e_2)\mapsto (-e_1,e_2)$.
$G$ has five irreducible representations, which we denote by $R_1, \cdots, R_5$, of dimension 1, 1, 1, 1, and 2, respectively. Here $R_1$ is the trivial representation, $R_2$ involves trivial action of $a$, $R_3$ involves trivial action of $x$, and $R_4$ involves trivial action of the diagonal reflection $ax$. The linear combinations of spin insertions that transform in each representation are 
\ie\label{r12345}
& R_1:~1,~ {s_0 (s_{e_1}+ s_{-e_1}+ s_{e_2}+ s_{-e_2})\over 4},~ {(s_{e_1} + s_{-e_1}) (s_{e_2} + s_{-e_2}) \over 4},~{s_{e_1}s_{-e_1}+s_{e_2}s_{-e_2}\over 2}, 
\\
&~~~~~~~~ s_{e_1}s_{-e_1} s_{e_2} s_{-e_2},~ {s_0s_{e_1} s_{-e_1} (s_{e_2} + s_{-e_2})+s_0s_{e_2} s_{-e_2} (s_{e_1} + s_{-e_1}) \over 4};
\\
& R_2:~{\rm none};
\\
& R_3: ~ {s_0 (s_{e_1}+ s_{-e_1}- s_{e_2}- s_{-e_2})\over 4}, ~ {s_{e_1}s_{-e_1}-s_{e_2}s_{-e_2}\over 2},
~ {s_0s_{e_1} s_{-e_1} (s_{e_2} + s_{-e_2}) - s_0s_{e_2} s_{-e_2} (s_{e_1} + s_{-e_1}) \over 4};
\\
& R_4:~ {(s_{e_1} - s_{-e_1}) (s_{e_2} - s_{-e_2}) \over 4} ;
\\
& R_5:~  {s_0 (s_{e_2} - s_{-e_2})\over 2},~ {(s_{e_1}+ s_{-e_1})( s_{e_2} - s_{-e_2} ) \over 4},
~ {s_0s_{e_1} s_{-e_1} (s_{e_2} - s_{-e_2}) \over 2}.
\fe
In the last row, we have listed only the $x$-invariant operator in each of the $R_5$ representations. The corresponding square positivity conditions are
\ie{}
& R_1: ~ \begin{pmatrix}1 & x_1 & x_3 & x_2 & x_4 & x_5 \\ x_1 & {1+x_2 + 2x_3\over 4} & {x_1+x_5\over 2} & {x_1+x_5\over 2} & x_5 & {x_2  +2x_3 + x_4\over 4} \\ x_3 & {x_1+x_5\over 2} & {1+ 2x_2 + x_4 \over 4} & x_3 & x_3 & {x_1 + x_5\over 2} \\ x_2 & {x_1+x_5\over 2} & x_3 & {1+x_4\over 2} & x_2 & {x_1+x_5\over 2} \\ x_4 & x_5 & x_3 & x_2 & 1 & x_1  \\  x_5 & {x_2 + 2x_3+x_4\over 4} & {x_1+x_5\over 2} & {x_1+x_5\over 2} & x_1 & {1+x_2+2x_3\over 4}  \end{pmatrix}\succeq 0, ~~~~~~R_4:~ {1- 2x_2+x_4\over 4}\geq 0,
\\
& R_3: ~ \begin{pmatrix} {1+x_2-2x_3\over 4} & {x_1 - x_5\over 2} & {2x_3 - x_2 - x_4\over 4} \\ {x_1 - x_5\over 2} &  {1-x_4\over 2} & {x_1-x_5\over 2}  \\ {2x_3 - x_2 - x_4\over 4} &  {x_1-x_5\over 2} & {1+x_2-2x_3\over 4} \end{pmatrix}\succeq 0,
~~~~~~R_5:~ \begin{pmatrix} {1-x_2\over 2} & {x_1 - x_5\over 2}  & {x_2 - x_4\over 2}  \\ {x_1-x_5\over 2}  & {1 - x_4 \over 4}  & {x_1 - x_5\over 2} \\ {x_2-x_4\over 2}  & {x_1 - x_5\over 2}  & {1-x_2\over 2}  \end{pmatrix}\succeq 0.
\fe

The first Griffiths inequality on $\cD$ amounts to
\ie
x_i\geq 0,~~~i=1,\cdots,5.
\fe
The second Griffiths inequality (\ref{secondgriff}) in the case $B=A^g$ for some $g\in G$ can be written equivalently as
\ie\label{aagineq}
\begin{pmatrix} 1 & \vv{\un{s}_A} \\ \vv{\un{s}_A} & \vv{\un{s}_A \un{s}_{A^g}} \end{pmatrix} \succeq 0.
\fe
It suffices to consider $A\subset \cD$ with $|A|=2,4$. Explicitly in terms of the $x_i$'s, the nontrivial instances of (\ref{aagineq}) amount to the positive semi-definiteness of the following matrices
\ie{}
& \begin{pmatrix} 1 & x_1 \\ x_1 & x_{2,3}  \end{pmatrix},~~ \begin{pmatrix} 1 & x_3 \\ x_3 & x_{2,4}  \end{pmatrix},~~\begin{pmatrix} 1 & x_2 \\ x_2 & x_4  \end{pmatrix},~~\begin{pmatrix} 1 & x_5 \\ x_5 & x_{2,3}  \end{pmatrix}.
\fe
Additional Griffiths inequalities of the form (\ref{secondgriff} within the 131 diamond where $A$ and $B$ are not related by symmetries are 
\ie\label{extragriff}
& x_5\geq x_1 x_2, x_1x_4, 
\\
& x_{2,3,4}\geq x_1 x_5,
\\
& x_1\geq x_3 x_5, x_2 x_5, x_4 x_5.
\fe
(\ref{extragriff}) cannot be represented equivalently in the form $\cM\succeq 0$ for a matrix $\cM$ whose entries are linear in the $x_i$'s, although one can write weaker inequalities of the latter form. In most of the examples we have examined so far, when the spin-flip equations and the reflection positivity conditions are imposed, the second Griffiths inequalities are automatically satisfied and thus need not be imposed as constraints. There is, however, an important class of exceptions, to be discussed in section \ref{sec:secondgriff}.

\subsection{Reflection positivity on the 3D cubic lattice}
\label{sec:3dreflect}

Reflection positivity conditions on the 3D cubic lattice, up to symmetries of the lattice, come in the same three types $(R_{e_1,0})$, $(R_{e_1, {1\over 2}})$, and $(R_{e_1+e_2,0})$, as described in (\ref{rxdef}), (\ref{twodrefl}). While the automorphism group of the lattice (fixing the origin) is the octahedral group, the reflection positivity conditions are organized according to symmetries of half of the lattice, which is the dihedral group in the case of $(R_{e_1,0})$, $(R_{e_1, {1\over 2}})$, and $\mathbb{Z}_2\times \mathbb{Z}_2$ in the case of $(R_{e_1+e_2,0})$.

Due to the limitation of computing power and efficiency of the algorithm implemented thus far, the largest domain $\cD$ we have considered in the cubic lattice is the 15551 region of Figure \ref{domains}, and the relevant reflection positivity conditions concerns linear combination of spin insertions within the domain $\cD$ on one side of the mirror plane. As the 15551 region is not invariant under all rotation symmetries of the lattice, there are 8 inequivalent sets of reflection positivity conditions, associated with the reflection map $R_{v,c}$ (\ref{rxdef}), with the following choices of the vector $v$ and shift $c$,
\ie{}
(v, c) = (e_1, 0),~ (e_1, {1\over 2}),~ (e_3, 0),~ (e_3, {1\over 2}),~(e_1+e_2, 0),~(e_1+e_2, 1),~(e_1+e_3, 0),~(e_1+e_3, 1).
\fe
For numerical SDP it will be sufficient to classify the reflection positivity conditions associated with $v=e_1, e_3, e_1+e_2$ and $c=0$ according to irreducible representations of the $\mathbb{Z}_2\times \mathbb{Z}_2$ symmetry only, leading to a total of $5+ 3\times 4 = 17$ positive-semidefinite matrices whose entries are linear in the spin correlators. The largest of these matrices is of size $2400\times 2400$, coming from $(v,c) = (e_1,0)$ in the trivial representation of $\mathbb{Z}_2\times \mathbb{Z}_2$.


\section{Bootstrap bounds}
\label{sec:allbounds}

In this section, we present bootstrap bounds obtained by imposing reflecton positivities and the first Griffiths inequality together with the spin-flip equations. First, we summarize how to set up the relevant SDP (\ref{sdpprob}). Once the region of the lattice $\cD$ is specified, we identify the relevant symmetry group for each of the reflections. Then, the positivity matrix $\cal M$, containing all the reflection positivity matrices, is written in terms of its blocks $X^{(k)}$ corresponding to irreducible representations $R_k$ of the symmetry group as discussed in section \ref{sec:invsdp}. We can express $X^{(k)}$ in terms of spin configurations of the primitive subsets $A\subset\cD$ (including the empty set) as
\ie
X^{(k)}=\sum_{A\subset\cD}Y^{(k)}_A\vv{\un{s}_A}\succeq 0,~~~ \forall k,
\fe
where $Y^{(k)}_A$ are the symmetric coefficient matrices. We then use the numerical solution of the spin-flip equations, discussed in section \ref{sec:loop}, to write $\vv{\un{s}_A}$ in terms of the independent variables $\vv{\un{s}_I}$ as $\vv{\un{s}_A}=\sum_I a^I_A\vv{\un{s}_I}+c_A$ with numerical coefficients $a^I_A$ and $c_A$, thus arriving at
\ie\label{positivityind}
X^{(k)}=\sum_{I}W^{(k)}_I\vv{\un{s}_I}+V^{(k)}\succeq 0, ~~~ \forall k,
\fe
where $W^{(k)}_I=\sum_A a^I_A Y^{(k)}_A$ and $V^{(k)}=\sum_Ac_AY_A^{(k)}$. In addition, the positivity matrix $\cal M$ includes $1\times1$ block matrices $\vv{\un{s}_A}$ of spin correlators for the primitive subsets, corresponding to the first Griffiths inequality. These can be written as
\ie\label{G1ind}
\vv{\un{s}_A}=\sum_I a^I_A\vv{\un{s}_I}+c_A\geq0,
\fe
for all $A$. The final form of the SDP problem (\ref{sdpprob}) is then given by
\ie\label{sdpind}
{}&\min_{y_I\in\mathbb{R}} \sum_I b^I y_I
\\
&\text{subject to }~\sum_Ia^I_Ay_I+c_A\geq0,~~\forall A,
\\
&~~~\text{and }~\sum_{I}W^{(k)}_Iy_I+V^{(k)}\succeq 0,~~ \forall k.
\fe
for a given $b^I\in\mathbb{R}$.

In order to solve (\ref{sdpind}), we used two SDP solvers: MOSEK \cite{mosek} and SDPA-QD \cite{5612693}. The main difference between the two solvers is their precision, where MOSEK is a double-precision solver while SDPA-QD is a quad-double precision solver. For this reason, even though MOSEK is much faster than SDPA-QD in cases where the SDP problem (\ref{sdpind}) can be solved by both,\footnote{For 2D Ising model on 13531 diamond with nonzero magnetic field $h$, a single run of MOSEK takes $\sim 30$ seconds to produce the solution on the 10-core Intel i9-10900F processor. In contrast, a single run of SDPA-QD takes $\sim 5$ hours to produce the solution on the Harvard RC Cluster. For SDPA-QD we used the parameters $\rm{betaStar}=0.3$, $\rm{betaBar} = 0.2$, $0.75 \le \rm{gammaStar} \le 0.9$, set the tolerance to be $\rm{epsilonStar} = 10^{-20}$ and the primal and dual feasibility error to be bounded by $\rm{epsilonDash} = 10^{-20}$ (in the notation of the SDPA manual \cite{citeSDPA}). The rest of the parameters were left at their default values.} it does not necessarily produce the optimal bounds within the relevant precision in some cases. In particular for some range of the couplings, the gap between the upper and lower bounds is of smaller order than the difference between the primal and dual solutions produced by MOSEK with default tolerance settings, and higher tolerance settings typically failed to produce solutions. Moreover, the performance of MOSEK depended heavily on the scales appearing in the coefficient matrices $W_I^{(k)}$ and $V^{(k)}$. When the maximum $M_{max}$ of the magnitudes of all the coefficients exceeded $\sim10^8$, MOSEK mostly failed to produce solutions and even when solutions were produced, they were sub-optimal. Therefore, in cases where the relevant precision is high or $M_{max}$ exceeded $\sim10^8$, we used SDPA-QD. We divided the coefficient matrices $W_I^{(k)}$ and $V^{(k)}$ by a common rescaling factor of the scale $\sim \sqrt{M_{max}}$ to increase the stability of the solvers.

\subsection{2D Ising at zero magnetic field}
\label{sec:2dhzero}

\begin{figure}[h!]
\centering
\includegraphics[width=1\textwidth]{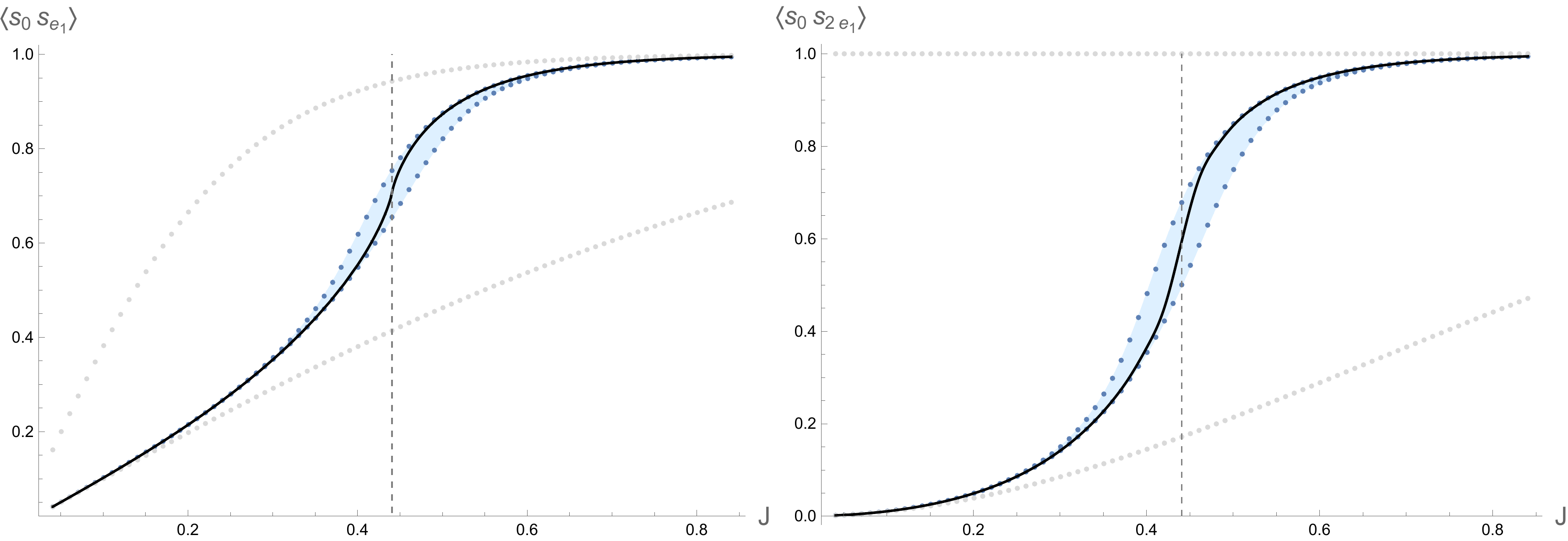}
\caption{The nearest-spin correlator $\vv{s_0 s_{e_1}}$ and next-to-nearest-spin correlator $\vv{s_0 s_{2e_1}}$ as functions of the coupling $J$ at $h=0$: the upper and lower bootstrap bounds obtained from the 131 diamond (gray dots) and the 13531 diamond (blue dots), compared to the exact result (black curve). The critical coupling $J_c = {\text{ln}(1+\sqrt{2})\over2}$ is marked with the vertical gray dashed line.}
\label{fig:h0S11}
\end{figure}

The 2D Ising model on the square lattice in the absence of magnetic field $h$ is exactly solvable (see \cite{McCoy:2012yc} for a review of the relevant formulae). In particular, the correlator of two spins separated by $N$ sites in one direction is given by
\ie
\vv{s_0 s_{N e_1}}_{h=0} = \det_{1\leq k,\ell\leq N} ( D_{k\ell} ),~~~~ D_{k\ell} = a_{k-\ell},
\fe
with\footnote{The branch of the square root in the integrand is chosen such that the latter is analytic as $e^{i\theta}$ moves around the unit circle.}
\ie
a_n = \int_0^{2\pi} {d\theta\over 2\pi} e^{-in\theta} \left[ { (1- e^{i\theta} e^{-2J} \tanh J ) (1- e^{-i\theta} e^{-2J} \coth J ) \over (1- e^{-i\theta} e^{-2J} \tanh J ) (1- e^{i\theta} e^{-2J} \coth J ) } \right]^{1\over 2}.
\fe
The comparison of the bootstrap bounds obtained from the spin-flip equation and reflection positivity condition on the 131 and 13531 diamonds (Figure \ref{domains}) against the analytic results for $\vv{s_0 s_{e_1}}$ and $\vv{s_0 s_{2 e_1}}$ at $h=0$ is shown in Figure \ref{fig:h0S11}.

\begin{figure}[h!]
	\centering
	\includegraphics[width=0.9\textwidth]{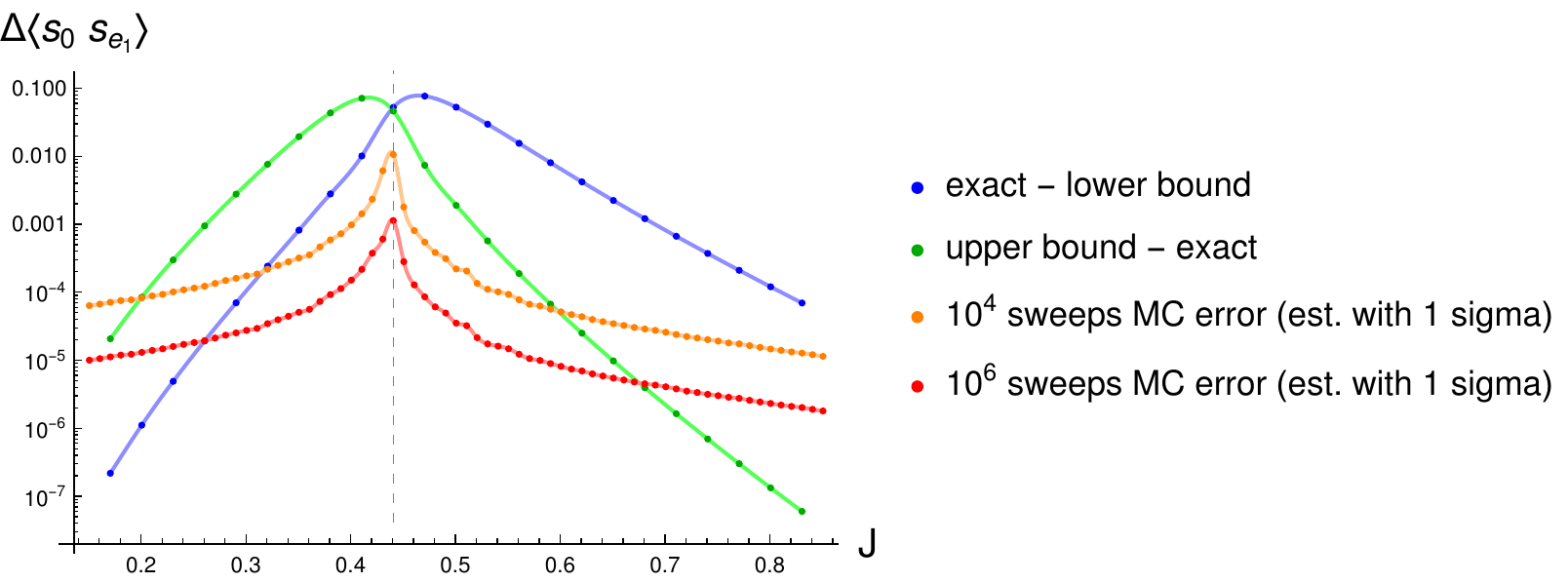}
	\caption{For the nearest-spin correlator $\vv{s_0 s_{e_1}}$ at $h=0$, the gaps between the upper/lower bootrstrap bounds and the exact value are shown as functions of $J$ on logarithmic scale, and compared to the estimated error bar of Monte Carlo simulation on a $200\times 200$ lattice with $\mathcal{O}(10^4)$ and $\mathcal{O}(10^6)$ sweeps of the Metropolis algorithm. }
	\label{fig:logplotS11}
\end{figure}

Here we can make several observations. First, the bootstrap bounds based on the 13531 diamond is dramatically stronger than that of the 131 diamond. Second, the window allowed by the two-sided bootstrap bounds (blue shaded region in Figure \ref{fig:h0S11}) is highly sensitive to the value of the coupling $J$, and narrows exponentially as $J$ moves away from the critical value $J_c={\ln (1+\sqrt{2})\over 2}$. Third, the upper bound is very close to the exact result for $J>J_c$, while the lower bound is close to the exact result for $J<J_c$. These features are further illustrated in Figure \ref{fig:logplotS11}, where the gap between the upper/lower bound and the exact result is shown in logarithmic scale, and compared to the typical error estimates of Monte Carlo simulations.\footnote{The Monte Carlo data was collected using the Metropolis algorithm and its error analysis following \cite{j._barkema_1999}. One ``sweep'' corresponds to one Metropolis step per lattice site.}

We can relax the assumption that the expectation values are invariant with respect to the $\mathbb{Z}_2$ symmetry that flips all spins, in which case the spin-flip equations (\ref{loopeqnab}) involve twice as many spin correlators. In a situation of spontaneous symmetry breaking, both the spin-flip equations and the positivity conditions hold for the correlators in a symmetry-breaking phase. One way to understand this is that one can select the phase by choosing a boundary condition on a finite lattice, say Dirichlet with all spins up at the boundary, where the spin-flip equations and positivity conditions still hold, and translation symmetry is restored in the limit that the lattice size is taken to infinity. Alternatively, one can view the correlators in the symmetry-breaking phase as the $h\to 0^+$ limit of the correlators in the presence of nonzero magnetic field, where the positivity conditions take an identical form to the $h=0$ case, and the spin-flip equations approach that of $h=0$ in the limit as well.

\begin{figure}[h!]
\centering
\includegraphics[width=0.75\textwidth]{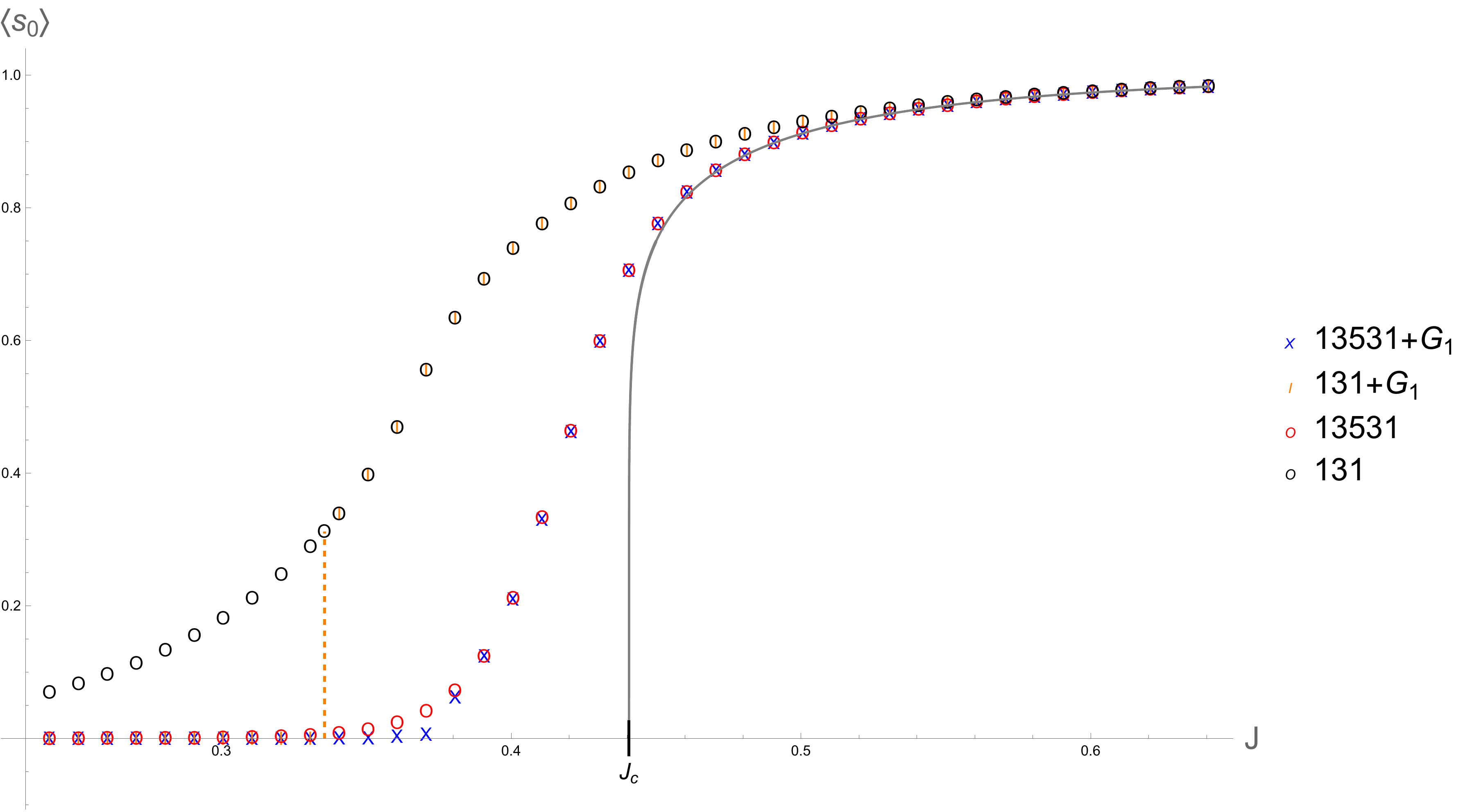}
\caption{Spontaneous magnetization $\vv{s_0}$ as a function of the coupling $J$ at $h=0$: upper bounds from the bootstrap obtained from the 131 diamond and the 13531 diamond, with and without the first Griffiths inequality $G_1$, in comparison to the exact result (gray curve). The orange vertical dashed line marks $J=\frac{\text{ln}(1 + 2 \sqrt{2})}{4}$ below which the upper bound from the 131 diamond with $G_1$ vanishes (see Appendix \ref{app:Gbound}).}
\label{fig:h0S1}
\end{figure}

The bootstrap upper bounds for $\vv{s_0}$, i.e. the spantanous magnetization, and their comparison to the known analytic result
\ie
\vv{s_0}_{h=0^+} = \left[ 1 - {1\over \sinh^4(2J)} \right]^{1\over 8},
\fe 
are shown in Figure \ref{fig:h0S1}. Interestingly, the first Griffiths inequality (denoted as $G_1$ in the figure) plays a nontrivial role here. The sharp transition from zero to nonzero upper bound on $\vv{s_0}$ at $J=\frac{\text{ln}(1 + 2 \sqrt{2})}{4}$ for the 131 diamond with $G_1$ (orange tick) is a result of the first Griffiths inequality as discussed in the Appendix \ref{app:Gbound}. When the first Griffiths inequality is not imposed, the upper bounds from both 131 and 13531 diamonds (red and black circles) are nonzero finite numbers. It is worth noting though that as the size of the diamond increases from 131 to 13531, the upper bounds approach the exact result even in the absence of the first Griffiths inequality.

 The bootstrap lower bounds for $\vv{s_0}$ are not displayed in Figure \ref{fig:h0S1} since they are always zero. This is because bootstrap always allows for the solution where any spin-flip odd configurations have zero expectation values due to the $\mathbb{Z}_2$ spin-flip symmetry, and the first Griffiths inequality states that these expectation values should be nonnegative. If the first Griffiths inequality is not imposed, the lower bounds on $\vv{s_0}$ differ from the upper bounds on $\vv{s_0}$ only by a sign.

\subsection{2D Ising at nonzero magnetic field}
\label{sec:2dnonzeroh}

\begin{figure}[h!]
\centering
\includegraphics[width=0.52\textwidth]{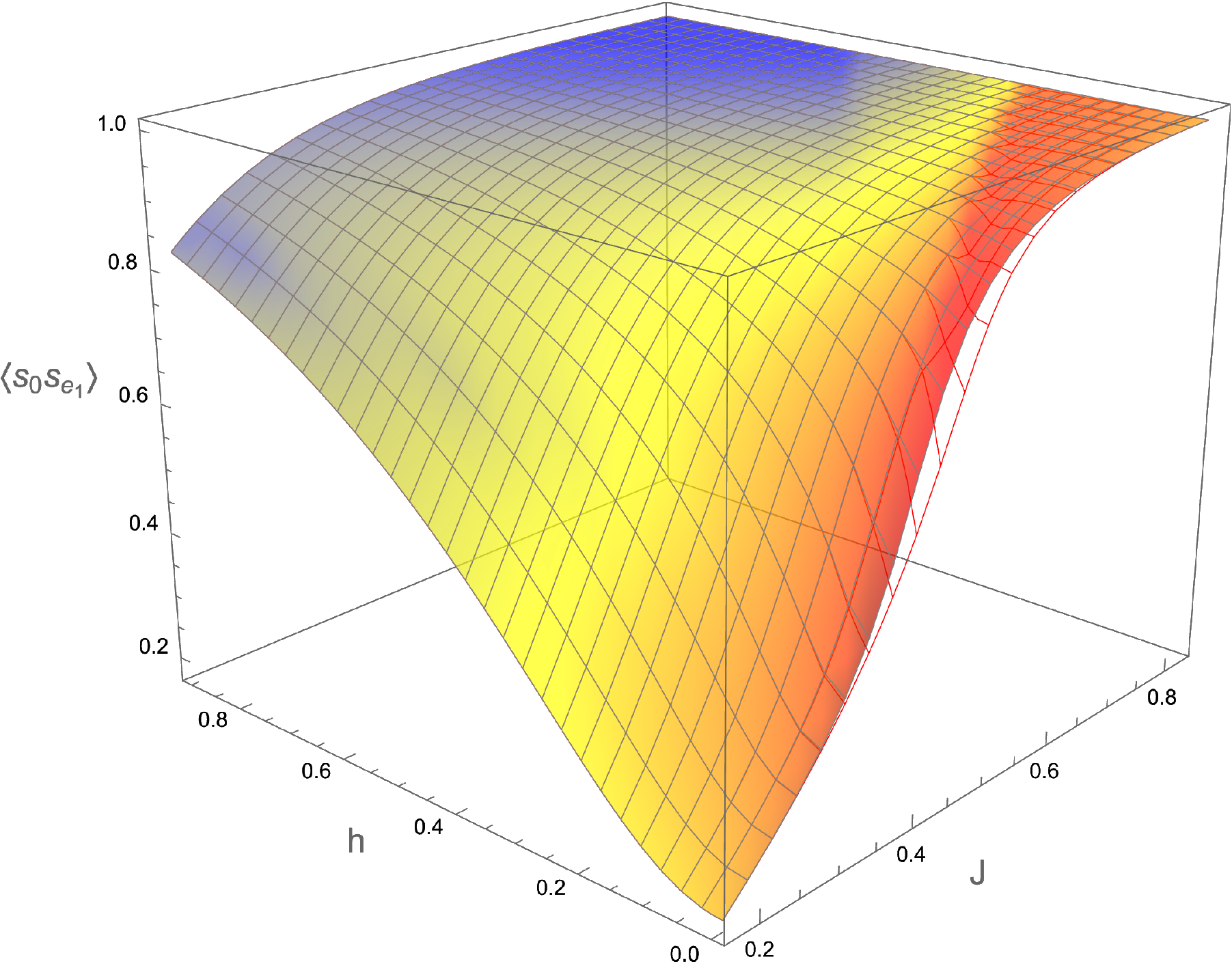}~~~~~~~\includegraphics[width=0.47\textwidth]{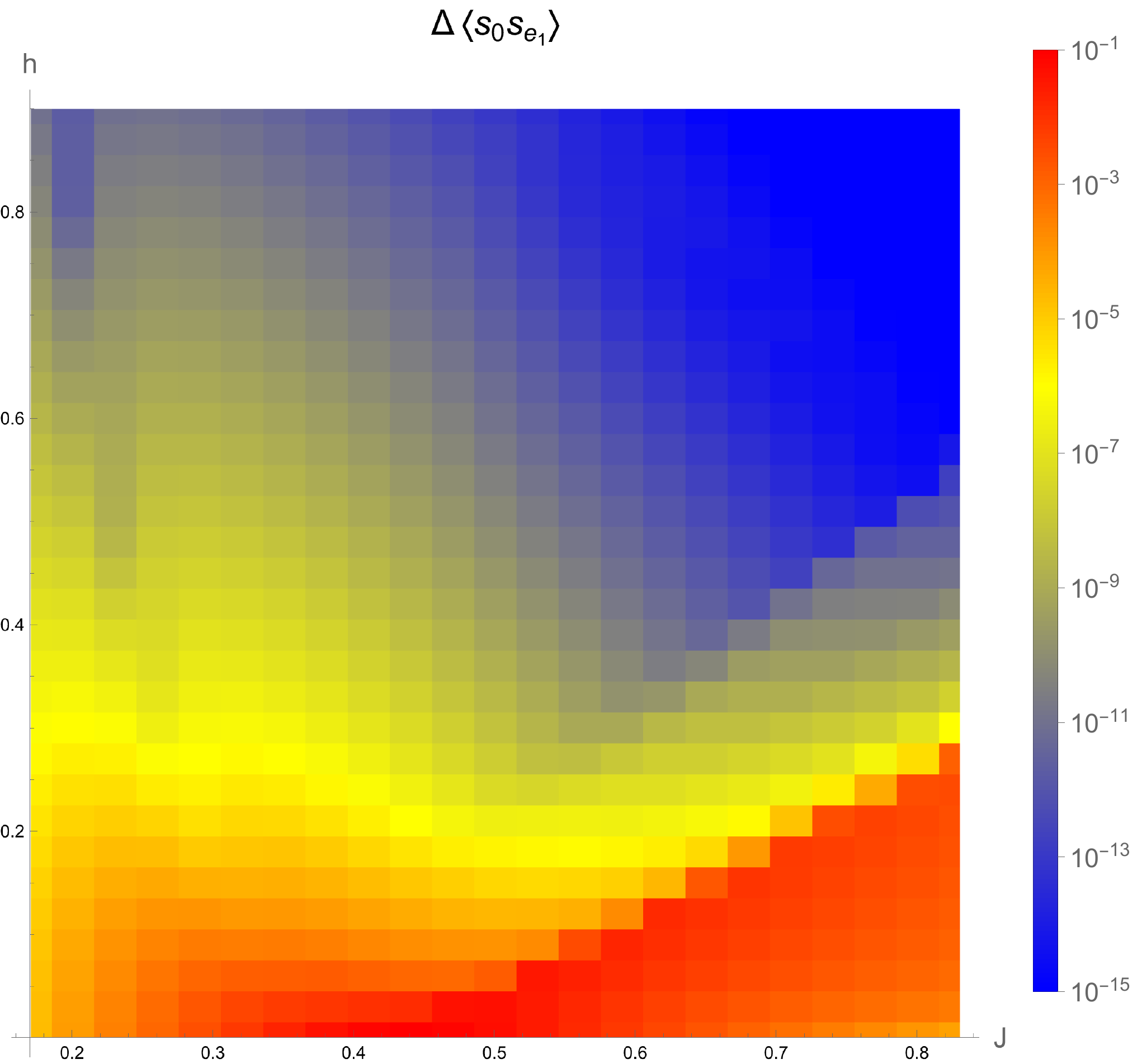}
\caption{Bootstrap bounds on $\vv{s_0s_{e_1}}$ as functions of the coupling $J$ and the magnetic field $h$, obtained by imposing reflection positivity and the first Griffiths inequality on the 13531 diamond. Left: upper bound is shown as the colored sheet with gray grids, while the lower bound is shown in red grids below. Right: the gap $\Delta{\vv{s_0s_{e_1}}}$ between the upper and lower bounds, in logarithmic scale, is represented by the color scheme ranging from red to yellow to blue.}
\label{fig:hS11}
\end{figure}

In the presence of nonzero magnetic field $h$, the 2D Ising model is not known to be exactly solvable. The bootstrap bounds for $\vv{s_0 s_{e_1}}$ based on spin-flip equations, reflection positivities, and the first Griffiths inequality on the 13531 diamond are shown as functions of $(J,h)$ in the left plot of Figure \ref{fig:hS11}. The gap between the upper and lower bootstrap bounds is highly sensitive to the value of $(J, h)$, and is shown on logarithmic color through the color gradient plot on the right of Figure \ref{fig:hS11}. 

\begin{figure}[h!]
	\centering
	\includegraphics[width=0.45\textwidth]{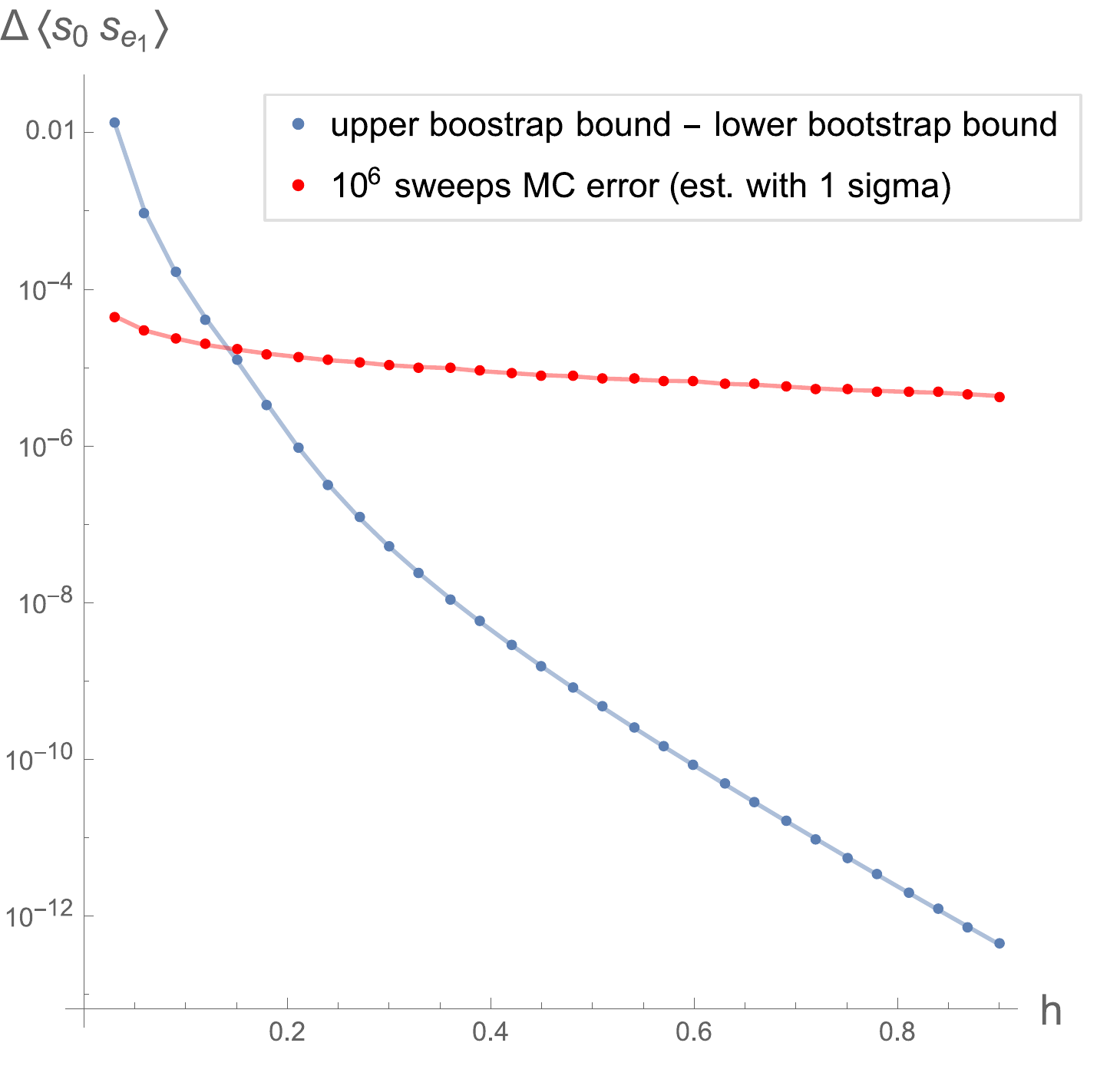}~~~~~~~~\includegraphics[width=0.45\textwidth]{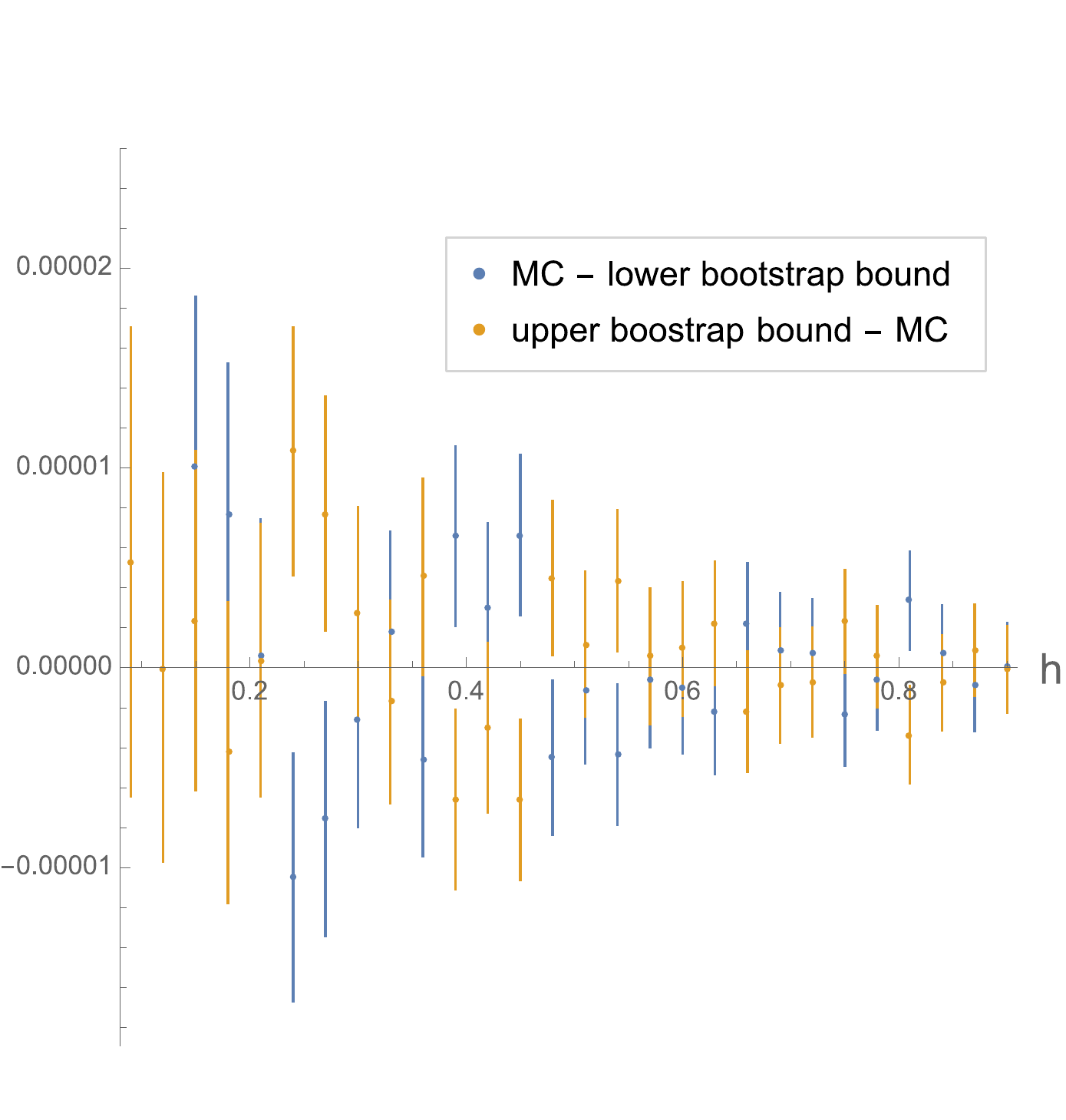}
	\caption{Left: the gap between upper and lower bootstrap bounds (blue dots) on the nearest-spin correlator $\vv{s_0 s_{e_1}}$ as a function of the coupling $h$ at $J=J_c$, obtained from the 13531 diamond, compared to the size of the error bar of Monte Carlo simulation (red dots) obtained on a $200\times 200$ lattice with $\mathcal{O}(10^6)$ Metropolis sweeps. Right: the difference between the bootstrap bounds and the MC data, where we have also displayed the 1 sigma error bars of the MC results. Note that the 2 sigma error bars would extend above the horizontal axis for all data points, indicating compatibility with the bootstrap bounds. }
	\label{fig:hS11err}
\end{figure}

The width of the window allowed by the bootstrap bounds, as a function of $h$ along the slice $J=J_c$, is shown on logarithmic scale in Figure \ref{fig:hS11err}, and compared to the typical error bar attainable with Monte Carlo simulation (based on $200\times 200$ lattice). While the bootstrap bounds obtained so far are somewhat loose near the critical point, it dramatically tightens with increasing $h$. For $h\gtrsim 0.2$, the bootstrap bounds are far more restrictive than MC results. We emphasize that the bootstrap bounds are rigorous results for the strictly infinite lattice.\footnote{One can verify, for instance, that the exact result for $\langle s_0 s_{e_1} \rangle$ on a $5\times 5$ periodic lattice at $h=0.3$, where the correlation length is sufficiently short that the finite size correction is less than $10^{-4}$, lies outside of the bootstrap bounds.} Note that the bootstrap bounds on $\vv{s_0}$  and $\vv{s_0 s_{e_1}}$ rule out the MC results at 1 sigma in some cases, but are compatible at 3 sigma in all cases we have examined (see right plot of Figure \ref{fig:hS11err} for a sample comparison).

\begin{figure}[h!]
	\centering
	\hspace*{-1.1cm}\includegraphics[width=1.0\textwidth]{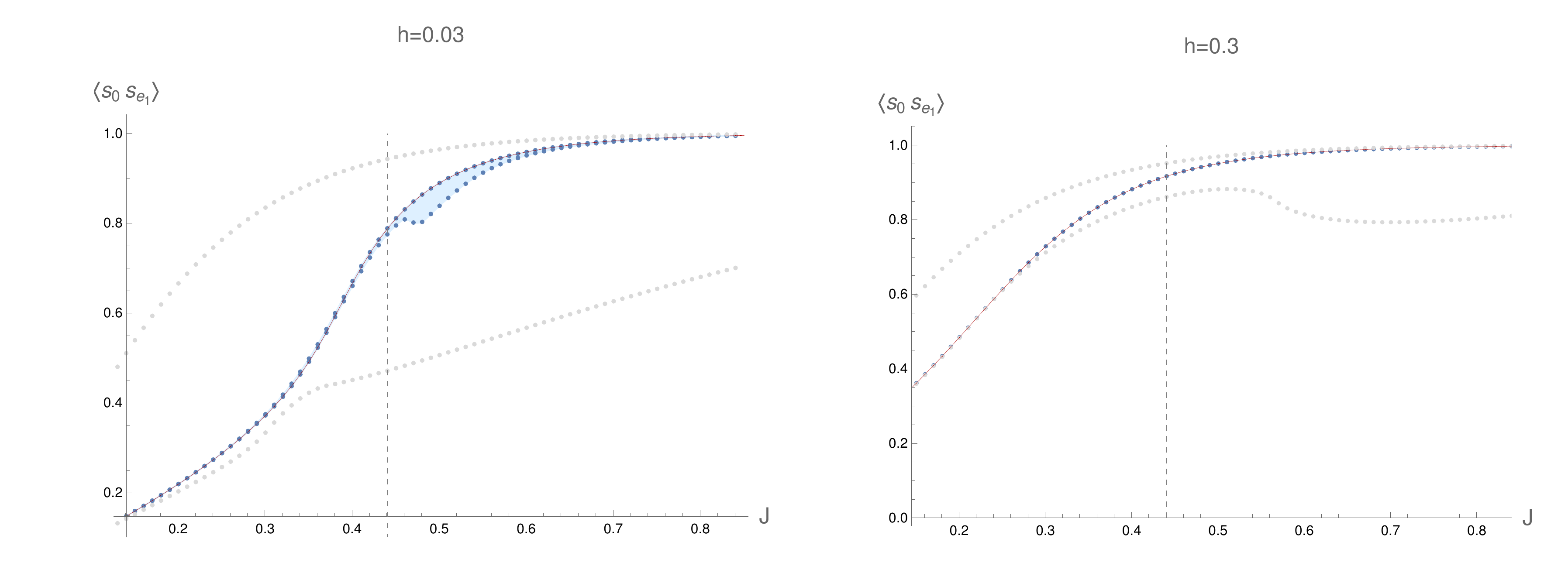}
	\hspace*{0cm}\includegraphics[width=1.0\textwidth]{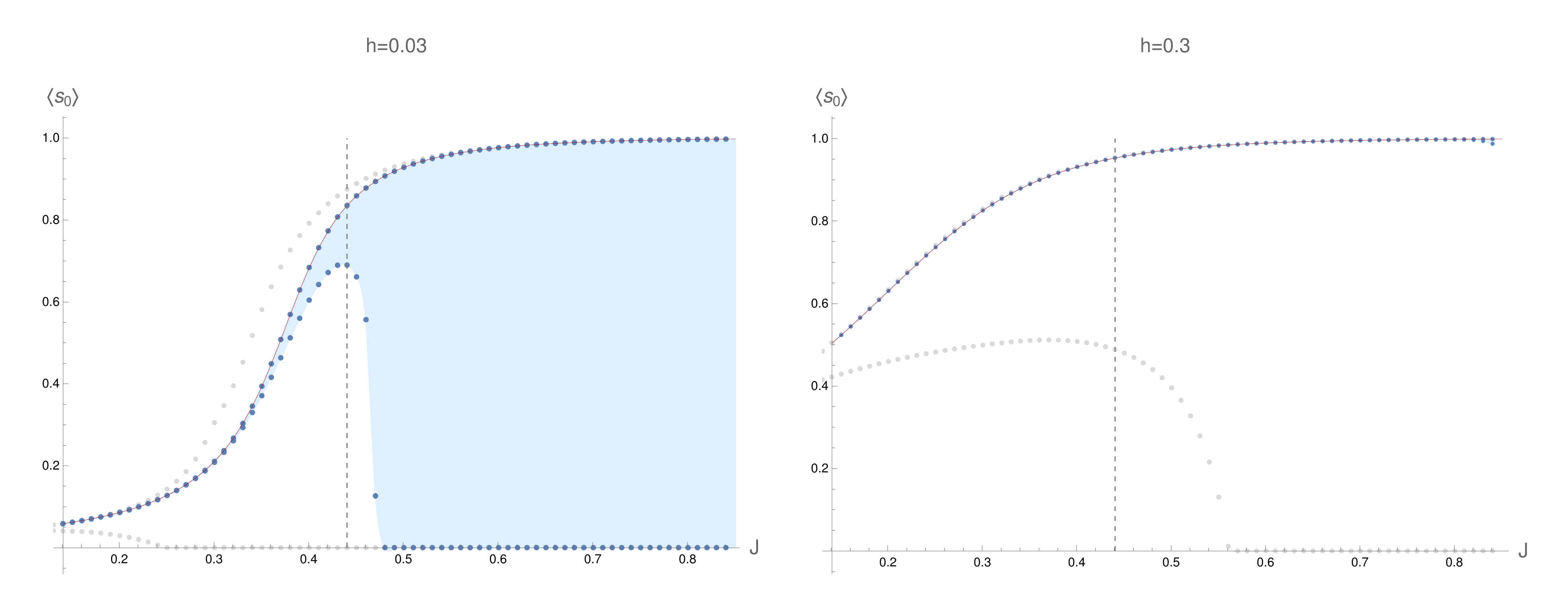}
	\caption{ The nearest-spin correlator $\vv{s_0 s_{e_1}}$ (top) and the magnetization $\vv{s_0}$ (bottom) as functions of $J$, at $h=0.03$ and $h=0.3$ respectively. The two-sided bootstrap bounds from the 131 diamond are shown in gray dots. The bounds from the 13531 diamond are shown in blue dots, with the allowed region shaded with light blue (invisible in the $h=0.3$ case). Note the non-monotonic feature of the lower bound. For comparison, Monte Carlo results are shown as the red curve. 
}
	\label{fig:hfixedS11}
\end{figure}

Another curious feature of the bootstrap bounds shown in Figure \ref{fig:hS11} is that the red band along which the bounds are weak extends away from the critical point to nonzero $h$. This might be due to a second length scale in the low temperature phase, namely the size of the critical droplet in the false vacuum \cite{Gawedzki1987},\footnote{We thank Slava Rychkov for explaining the role of this length scale.} coinciding with the size of the domain $\cD$. Note that the lower bound is in fact not monotonic in $J, h$, and shows a ``dip'' near the red band, as illustrated more clearly in the top plots of Figure \ref{fig:hfixedS11}. This is presumably because we have not taken into account the second Griffiths inequality as part of the positivity constraints, as we will discuss in section \ref{sec:secondgriff}.

The analogous bootstrap bounds on the magnetization $\vv{s_0}$ at two nonzero values of the magnetic field $h$ are shown in the bottom plots of Figure \ref{fig:hfixedS11}. In contrast to the $h=0$ case (Figure \ref{fig:h0S1}), the lower bound is nontrivial and is expected to close in on the exact result with increasing size of the domain $\cD$, albeit more slowly at small $h$.

\subsection{The role of Griffiths inequalities}
\label{sec:secondgriff}

\begin{figure}[h!]
\centering
\includegraphics[width=1\textwidth]{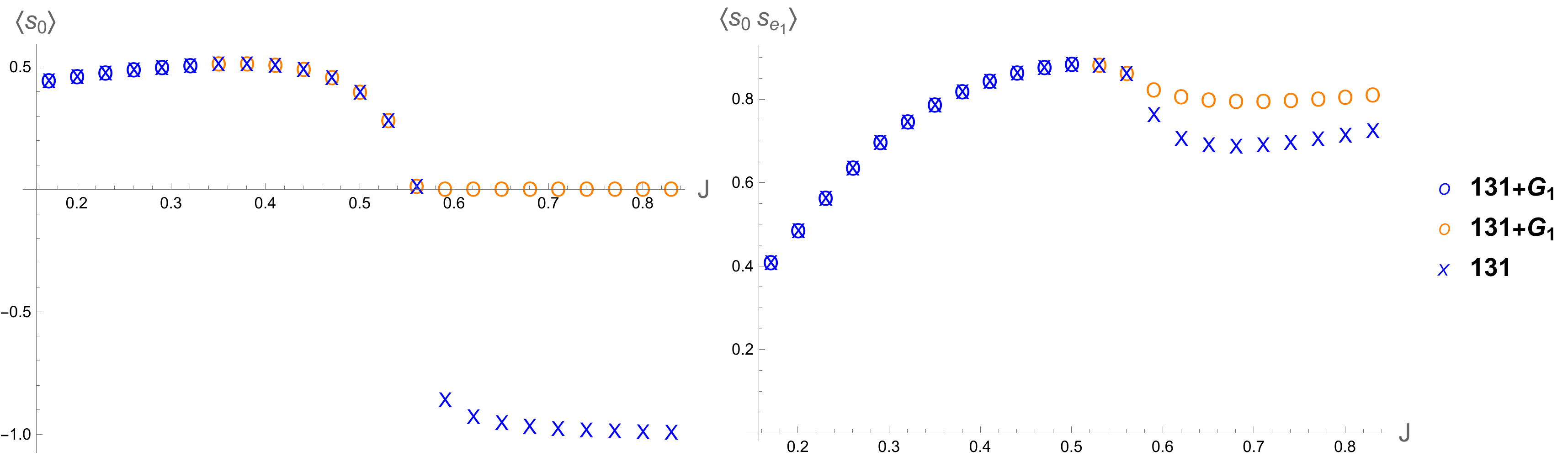}
\caption{ Lower bounds on $\vv{s_0}$ and $\vv{s_0 s_{e_1}}$ from 131 diamond at $h=0.3$, with (blue and orange circles) and without (blue crosses) the first Griffiths inequality $G_1$. Among the bounds with $G_1$, the second Griffiths inequality $G_2$ is violated for orange circles.}
\label{fig:hG12}
\end{figure}

So far we have only taken into account the first Griffiths inequality ($G_1$) (\ref{firstgriff}) in obtaining the SDP bounds. Indeed, $G_1$ is seen to improve the lower bound on $\vv{s_0}$ and $\vv{s_0 s_{e_1}}$, as illustrated in Figure \ref{fig:hG12}.

The second Griffiths inequality ($G_2$) (\ref{secondgriff}), on the other hand, cannot be straightforwardly implemented in SDP, except for the symmetric case considered in (\ref{aagineq}). It can be verified that the symmetric $G_2$ inequalities (\ref{aagineq}) do not improve the bounds shown in Figure \ref{fig:hG12}. However, the asymmetric $G_2$ inequalities such as (\ref{extragriff}) are expected to improve the bounds, as they are in fact violated by the solution to the SDP problem of minimizing $\vv{s_0}$ and $\vv{s_0 s_{e_1}}$ at the points represented by orange circles in Figure \ref{fig:hG12}.

As an example, consider $(J=J_c-0.09, h=0.3)$ corresponding to the left-most orange circle in the left plot of Figure \ref{fig:hG12}. Let us denote by $\vv{\cdots}_*$ the putative values of the correlators given by the solution to the SDP problem of minimizing $\vv{s_0}$. The following asymmetric $G_2$ inequality, together with several others, are explicitly violated:
\ie
\vv{s_{e_2}}_* - \vv{s_{-e_1}s_{e_1}s_{e_2}}_* \vv{s_{-e_1}s_{e_1}}_* =-0.00564... \ngeq 0.
\fe
The same inequality is also violated at the left-most orange circle in the right plot of Figure \ref{fig:hG12} at $(J=J_c+0.09, h=0.3)$.

A consequence of the $G_2$ inequalities is that every spin correlator $\vv{ \un{s}_A }$ is monotonically non-decreasing as a function of $J$ and $h$ \cite{Glimm:1987ng}. One may expect that imposing the full set of $G_2$ inequalities will lead to monotonic lower bounds on the spin correlators as well. Indeed, the left-most orange circles on Figure \ref{fig:hG12} appear roughly around the maximum of the lower bounds, in agreement with this expectation. However, the $G_2$ inequalities are generally not convex and are difficult to implement systematically at present.

\subsection{3D Ising at zero magnetic field}
\label{sec:3dbounds}

\begin{figure}[h!]
\centering
\includegraphics[width=.6\textwidth]{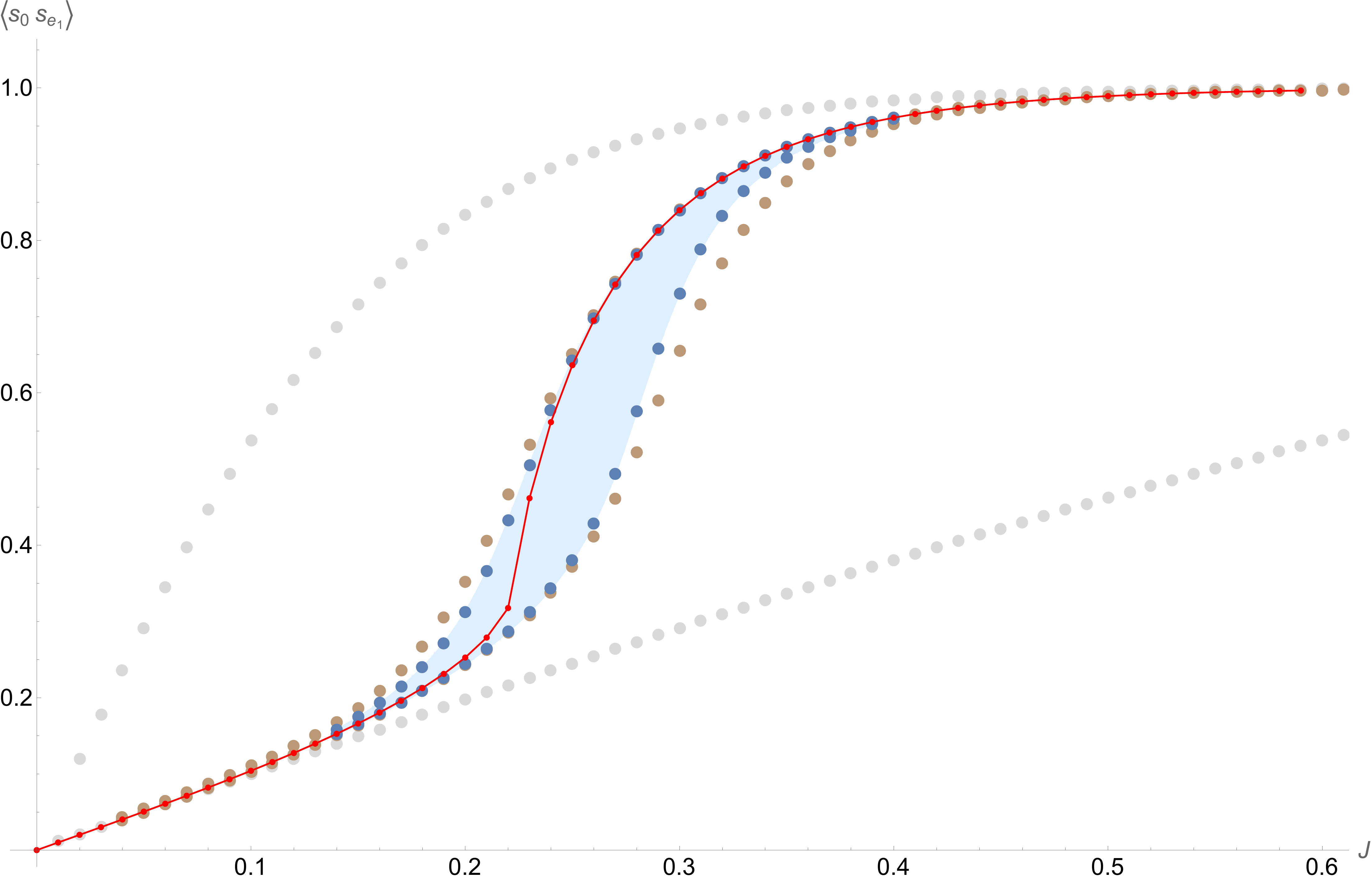}
\caption{The nearest-spin correlator $\vv{ s_0 s_{e_1} }$ in the 3D Ising model (at $h=0$) as a function of the coupling $J$: the upper and lower bootstrap bounds obtained from the 151 diamond (gray dots), the 1551 region
(brown dots), and the 15551 region with truncated reflection positivity matrices (blue dots), compared to results of Monte Carlo simulation (red dots, connected for visualization). The MC data were obtained on a $100^3$ lattice with $\mathcal{O}(10^4)$ Metropolis sweeps. }
\label{fig:3dS11bounds}
\end{figure}

For the 3D Ising model on the cubic lattice at $h=0$, we have considered spin-flip equations and reflection positivity conditions on regions up to the 15551 domain of Figure \ref{domains}. The substantial number of independent spin correlators (see section \ref{sec:algloop}) and size of the reflection positivity matrices have proved to be challenging for the SDP solvers we used so far. As a workaround, we simply truncated the reflection positivity matrices described in section \ref{sec:3dreflect} to their leading $100\times 100$ principal submatrices, which can be handled numerically with MOSEK and give less-than-optimal but nonetheless rigorous bounds. The resulting two-sided bounds on the nearest-spin correlator $\vv{ s_0 s_{e_1} }$, together with the analogous bounds obtained from the 1551 domain (with the full reflection positivity condition taken into account), is shown in Figure \ref{fig:3dS11bounds} and seen to be compatible with results of Monte Carlo simulation. 

\section{Discussion}
\label{sec:discuss}

The main result of this paper is that the spin-flip equations together with positivity conditions in a finite domain $\cD$ put {\it rigorous} two-sided bounds on spin correlators of Ising model on the {\it infinite} lattice. These bounds are numerically optimized through semidefinite programming, and narrow down the spin correlators dramatically with increasing size of the domain $\cD$. For 2D Ising model on the square lattice, with or without a background magnetic field, the bootstrap bounds based on the 13531 diamond (consisting of 13 sites) determine the energy density expectation value within a window that narrows exponentially as one moves away from the critical coupling. For a certain range of the coupling and magnetic field, the bootstrap bounds are tighter than the precision attainable with simple Monte Carlo methods (based on the Metropolis algorithm).

For 3D Ising model on the cubic lattice, the analogous next-to-simplest diamond-shaped region that is invariant under the octahedral automorphism group would be the ``1-5-13-5-1'' region, consisting of 25 sites. To solve the spin-flip identities for all spin correlators on this domain and to implement reflection positivity via SDP are beyond reach with our current algorithm. The results we have obtained (Figure \ref{fig:3dS11bounds}) based on smaller domains consisting of 12 or 17 sites, and a subset of the reflection positivity conditions in the latter case, are encouraging although not yet competitive against Monte Carlo methods in terms of determining the value of spin correlators in the parameter regime of interest. However, we would like to emphasize that the bootstrap bounds are rigorous, and that the upper/lower bounds are seen to be quite close to Monte Carlo results at coupling $J$ above/below the critical value $J_c$.

It is perhaps unsurprising that the bootstrap bounds are weakest near criticality, where correlation length grows while we have only taken into account spin-flip identities and positivity conditions on a fixed finite domain. Nonetheless, our results thus far suggest the following

{\it Conjecture:} the spin-flip equations (\ref{loopeqnab}) together with the reflection positivity condition\footnote{While the Griffiths inequalities are seen to improve the bootstrap bounds in certain cases, empirical evidence (e.g. Figure \ref{fig:h0S1} and \ref{fig:hfixedS11}) suggests that they are inessential for the convergence of the bounds with increasing domain.} (\ref{reflpos}) with (\ref{rxdef}), (\ref{twodrefl}) on a finite domain $\cD$ of the lattice lead to two-sided bounds on every spin correlator $\vv{\un{s}_A}$ that converge to the exact value as the size of $\cD$ increases to infinity when there is a single phase. The same assertion applies to $\mathbb{Z}_2$-invariant spin correlators when there are two phases related by the $\mathbb{Z}_2$-symmetry, i.e. for $h=0$ and $J>J_c$.


While it would clearly be of interest to prove this conjecture, a more relevant question in practice is the rate of convergence of the bounds on $\vv{\un{s}_A}$ with the size of the domain $\cD$, which is expected to be highly sensitive to the size of the set of lattice sites $A$ in the correlator of interest, and the deviation of the coupling $J$ from criticality. To access long distance correlators near criticality, thereby probing observables of the continuum Ising field theory, is particularly challenging in our lattice bootstrap approach, and requires substantial improvement of the algorithm adopted in this work.

Indeed, there is vast room for improvement in terms of the efficiency of each step of our numerical bootstrap algorithm. First, it is conceivable that, to access long distance spin correlators of a few (e.g. a pair of) spins, only certain types of spin configurations need to be taken into account in the spin-flip equations, such as ones that involve a string of sites connecting the spin operator insertions in the correlator of interest. Second, the spin-flip equations formulated in this paper are excessively overcomplete. It should be possible to identify the linearly independent spin-flip equations a priori, rather than numerically as we have done so far. Finally, while the size of the positive-semidefinite matrices of spin correlators that arise in the reflection positivity condition grows quickly with the domain $\cD$, it is likely that a substantial truncation to principal submatrices would be sufficient for producing a two-sided bound whose width is of the same order as the optimal one. These improvements are left to future work.

\section*{Acknowledgements}

We would like to thank Mohamed Ali Belabbas, Jixun K. Ding, Sean Hartnoll, Vladimir Kazakov, Jacques H. H. Perk, Jiaxin Qiao, Slava Rychkov, and Zechuan Zheng for discussions and correspondences. We thank the organizers of Bootstrap 2022 in Porto, Portugal, for their hospitality during the course of this work. This work is supported in part by a Simons Investigator Award from the Simons Foundation, by the Simons Collaboration Grant on the Non-Perturbative Bootstrap, and by DOE grant DE-SC0007870. The computations in this work are performed on the FAS Research Computing cluster at Harvard University.

\appendix

\section{Coefficients in the spin-flip equations}
\label{sec:loopcoeff}

In the $d=2$ case, the nonzero $C_{n,\ell}$ coefficients appearing in (\ref{wzrel}) are\footnote{These formulae can be found using ``FindSequenceFunction'' in Mathematica. }
\ie{}
& C_{2k, 2} = {2^{2k} - 2^{4k-4} \over 3},~~~~ C_{2k,4} = {2^{4k-6} - 2^{2k-4} \over 3},
\\
& C_{2k+1,1} = {2^{2k+2} - 2^{4k}\over 3},~~~~ C_{2k+1,3} = {2^{4k-2}-2^{2k-2}\over 3}.
\fe
We can perform the infinite sums in (\ref{abforms}) to obtain the following non-vanishing $A_\ell, B_\ell$ coefficients,
\ie\label{abdtwocase}
& A_2 =  {-15 + 16 \cosh(4 J) - \cosh(8 J) \over 48},~~~~ A_4 = {3 - 4 \cosh(4 J) + \cosh(8 J) \over 192},
\\
& B_1 = {  8 \sinh(4 J) - \sinh(8 J) \over 12},~~~~ B_3 = {-2\sinh(4 J) + \sinh(8 J)\over 48}.
\fe

In the $d=3$ cases, the nonzero $C_{n,\ell}$'s are
\ie{}
& C_{2k,2} = C_{2k-1,1} = -{ 2^{-4 + 2 k} (-270 + 27 \cdot 2^{2 k} - 2 \cdot 9^k) \over 45},
\\
& C_{2k,4}= C_{2k-1,3}  = { 2^{-7 + 2 k} (-39 + 3 \cdot 2^{2 + 2 k} - 9^k) \over 9},
\\
& C_{2k,6}= C_{2k-1,5} = - { 2^{-9 + 2 k} (-15 + 3 \cdot 2^{1 + 2 k} - 9^k) \over 45}.
\fe
After performing the sums, we obtain
\ie{}
& A_2 = {-245 + 270 \cosh(4 J) - 27 \cosh(8 J) + 2 \cosh(12 J)\over 720},
\\
& A_4 = {28 - 39 \cosh(4 J) + 12 \cosh(8 J) - \cosh(12 J)\over 1152},
\\
& A_6 = { -10 + 15 \cosh(4 J) - 6 \cosh(8 J) + \cosh(12 J) \over 23040},
\\
& B_1 = { 45 \sinh(4 J) - 9 \sinh(8 J) + \sinh(12 J) \over 60},
\\
& B_3 = {-13 \sinh(4 J) + 8 \sinh(8 J) - \sinh(12 J) \over 192},
\\
& B_5 = { 5 \sinh(4 J) - 4 \sinh(8 J) + \sinh(12 J)\over 3840}.
\fe

\section{A sample analytic bootstrap bound}\label{app:Gbound}

A simple analytic bound on the spin magnetization can be obtained by consideration of the 131 diamond described in section \ref{sec:2dexloop}. In addition to the spin correlators (\ref{xcorrs}), we have spin correlators with odd number of spins,
\ie
y_1 = \langle s_0 \rangle, ~~~ y_2 = \langle s_{-e_1} s_0 s_{e_1} \rangle, ~~~ y_3 = \langle s_0 s_{e_1} s_{e_2} \rangle, \\
y_4  = \langle s_{-e_1} s_{e_1} s_{e_2} \rangle, ~~~ y_5 = \langle s_{-e_1} s_0 s_{e_1} s_{-e_2} s_{e_2} \rangle.
\fe
The additional spin-flip equations, corresponding to spin configurations $s_A$ with $A = \{ 0 \}$, $\{ 0, \pm e_1 \}$, $\{ 0, e_1, e_2 \}$, $\{ \pm e_1, e_2 \}$, and  $\{ 0, \pm e_1, \pm e_2 \}$, determine all correlators but $y_1$ to be
\ie
y_2 = y_3 & = - y_1 \frac{1 + \cosh(4 J) - 2 \sinh(4 J)}{2\sinh^2(2 J)}  ,\\
y_4 & = - y_1 \frac{ 2 \cosh(2 J) + 2 \cosh(6 J) + \sinh(2 J) - 3 \sinh(6 J)) }{4\sinh^3(2 J)}, \\
y_5 & =  - y_1 \frac{1 + 3\cosh(4 J) - 4 \sinh(4 J)}{2\sinh^2(2 J)} .
\fe
Assuming $J,y_1 \geq 0$, each first Griffiths inequality $y_i\geq 0$, with $i=2,4,5$, therefore implies a rigorous bound on the magnetization, the strictest coming from $y_5 \geq 0$:
\ie
y_1 = 0 ~~ \text{for } J \leq \frac{\text{ln}(1 + 2 \sqrt{2})}{4}.
\fe

\bibliographystyle{JHEP}
\bibliography{Ising}

\end{document}